\documentclass[a4paper,11pt]{article}
\usepackage[latin1]{inputenc}
\usepackage{indentfirst}

\usepackage{amsfonts}
\usepackage{float}
\usepackage{latexsym,amssymb,amsmath}
\usepackage[dvips]{graphics,graphicx,epsfig,color}
\usepackage[lofdepth,lotdepth]{subfig}
\begin{document}
\thispagestyle{empty}
\title{\bf On the velocity of sound in water: theoretical aspects of Colladon's nineteenth century experiments}
\author{\bf Armand Wirgin \thanks {LMA, CNRS, UPR 7051, Aix-Marseille Univ, Centrale Marseille, F-13402 Marseille Cedex 20, France}}
\date{\today}
\maketitle
\begin{abstract}
In 1827, Colladon carried out a series of experiments
in Lac Leman (Lake Geneva, Switzerland) to measure the speed of sound in water.
 The purpose of our contribution is to treat this measurement as an inverse problem, and show, by theory how to solve the latter. It is thus revealed under what circumstances it is legitimate to employ the time-of-flight scheme underlying the Colladon experiments and how to bypass this scheme in order to fully  account for  the temporal and geometric characteristics of the source (of sound), the temporal characteristics of the received signal and the error incurred by the finite distance between the source and receiver.
\end{abstract}
%%%%%%%%%%%%%%%%%%%%%%%%%%%%%%%%%%%%%%%%%%%%%%%%%%%%%%%%%%%%%%%%%%%%%%%
\newpage
%%%%%%%%%%%%%%%%%%%%%%%%%%%%%%%%%%%%%%%%%%%%%%%%%%%%%%%%%%%%%%%%%%%%%%%%
\tableofcontents
%%%%%%%%%%%%%%%%%%%%%%%%%%%%%%%%%%%%%%%%%%%%%%%%%%%%%%%%%%%%%%%%%%%%%%%%
\newpage
%%%%%%%%%%%%%%%%%%%%%%%%%%%%%%%%%%%%%%%%%%%%%%%%%%%%%%%%%%%%%%%%%%%%%%%%
\section{Introduction}
The measurement of the bulk velocity of sound in  various kinds  of homogeneous media (solid, liquid, gaseous) has been a subject of research for centuries (Baskevitch, 2008). It became quickly known (the non-infinitesimal time delay between a vocalized sound and its echo  in a mountainous environment being a readily-measurable experience to anybody)  that sound propagates with a velocity that is finite  (contrary to that of light which was at first thought to be infinite, and later to be  'very large'), and therefore susceptible of being measured by quite simple means (henceforth termed the kinematic method), usually deriving from the inversion of the formula $\Delta x=v\Delta t$, $v$ the velocity of sound, $\Delta x$ the known distance between two points (usually the point of sound emission and another point (of reception),  more or less far from the first point, and $\Delta t$ the time it takes for the sound to propagate from the first to second points. In this formula, $\Delta x$ and $\Delta t$ are observed quantities, from which $v$ is retrieved by inversion via the formula $v=\Delta x/\Delta t$. The same formula also reveals: i) the fact that if it is possible to observe only rather large $\Delta t$ then $\Delta x$ must be chosen to be correspondingly large, ii) the accuracy of the retrieval of $v$ depends heavily on the accuracy of the observation of $\Delta x$ which can be low for large $\Delta x$, iii) likewise, the accuracy of the retrieval of $v$ depends on the accuracy of the observation of $\Delta t$ which can be low for small $\Delta t$ (whence the interest in disposing of reliable means (e.g., clocks, watches, etc. (Derham, 1696; Derham, 1708)) for measuring time.   Finally, it must be stressed  that $v$ cannot be observed directly (contrary to $\Delta x$ and $\Delta t$, but rather that its measurement  is an inverse problem employing data related to a distance and interval of time.

There exists another approach to the problem of the measurement of $v$, namely from first (physical) principles. The first to do this was Newton who related the bulk velocity of sound to the density and compressibility of the material in which it propagates. This is again an inverse problem in which the data  is now connected to  observations of density and compressibility.

Newton found that the $v$ in air retrieved by the two methods differed considerably and all his rather contrived attempts to account for this difference failed. (Bruce, 2012; Westfall, 1973) This opened up a new avenue of research on the validity of Newton's formula, focused mainly on the velocity of sound in air. It led to Laplace's finding (Finn,1964; Roberts,2008) that Newton's formula takes no account of the influence of the heat generated in the air by the compression of this gas.

While most of the research (both theoretical and experimental) on the velocity of sound centered on that of air (Lenihan,1952) and other gases, Colladon and Sturm (Colladon \& Sturm, 1827a; Colladon \& Sturm, 1827b; Colladon \& Sturm, 1827c; Colladon, 1837; Colladon, 1893) chose to focus their attention on liquids, the most common of which is water. They did this firstly by the physical method based on data observed (by them) connected with density, compressibility and the heat generated during compression of the fluid, and secondly (after concluding that Newton's formula was correct in pure water since the heat problem is of negligible influence on the propagation of sound therein) by the kinematic method.

The remarkable work of Colladon and Sturm is of experimental nature, with much attention given to detail, especially in their experiments related to the kinematic method of specific interest herein. Although they made many efforts to eliminate all sources of error (in the kinematic method), it appeared to us that they were either  not aware, or underestimated the importance, of two problems having to do with the spectrum and geometric support of the source of pulse-like sound they propagated in the water and the influence of observational error concerning the onset time of the sound pulse. Taking into account these factors, as well as others concerning the influence of the far-field and paraxial region assumptions on retrieval error, will be the subject of the second part of this investigation.

Colladon \& Sturm'x publication was written in french and therefore not known to a large part of the scientific community. This is unfortunate since their work gives a good glimpse of some of the key issues that physicists wrestled with during the 17th to 19th centuries. Other than its historical interest, the work of Colladon \& Sturm has a sort of esthetic value that derives from the considerable ingeniosity employed in devising and carrying out the experiments.

Thus, we thought it useful to make a translation (into english) of Colladon and Sturm's paper so as to make it available to as large a  part as possible of the audience interested in acoustics and/or the history of science. Actually, a translation into german is available in (Colladon \& Sturm, 1828) and some interesting additions can be found in Colladon's, autobiography (Colladon, 1893), the english translation of which, (actually, only excerpts) are published in (Lindsay, 1973) as well as  in the  article (in german) (K\"oltzsch, 2011)  in this journal.

 Hereafter, my personal additions to published 19th century published papers will appear in square brackets (i.e., [...]).
%%%%%%%%%%%%%%%%%%%%%%%%%%%%%%%%%%%%%%%%%%%%%%%%%%%%%%%%%%%%%%%%%%%%%%%%
\subsection{Beudant and Colladon}
 In his textbook on physics (Beudant, 1824); section Liquids in motion, subsection Vibratory motion), Beudant alludes very vaguely to his experiments on the speed of sound in water. It is probable that Colladon read this book and later interrogated Beudant on details of his experiment. Thus, the Beudant experiments can reasonably be thought to have inspired Colladon's own 1826 experiments in Lake Leman.\\
%%%%%%%%%%%%%%%%%%%%%%%%%%%%%%%%%%%%%%%%%%%%%%%%%%%%%%%%%%%%%%%%%%%%%%%%
\subsubsection{Beudant's book}
On page 296 of his book (Beudant, 1824), Beudant writes;\\\\
Liquids are able to vibrate when they are in contact with a  body to which such motion has been procured. In effect, if one places water in a standing glass, and he passes his finger on the edge of the said glass to produce a sound, one immediately observes that the surface of the liquid is marked by wrinkles that move from the circumference to the center.

The vibratory property of liquids is also proven by the ease with which they propagate the sound produced at an arbitrary point in their volume. If, being submerged in water, one strikes a bell in the liquid, the intensity of the perceived sound will be much larger than if he was in air; the sound, by its intensity, is even very uncomfortable to the ear. One finds here another example of the providential goodness of the Creator, who did not give to the hearing organs of fishes all the development that can be observed in the majority of terrestrial animals.

The ability of transmitting sound is not the same in all liquids; it seems, from various experiments conducted in Turin by Mr. P\'erolle, that it is directly related to the specific weights [densities].

Mr. Laplace, by specific considerations, based on  the compression of which water is capable, gives the figure of 1526 m/s for the speed of sound in rainwater, and 1621 m/s in seawater. I already  thought to be able to determine the latter [seawater speed] to at least 1500 m/s, from experiments I made in the sea.
%%%%%%%%%%%%%%%%%%%%%%%%%%%%%%%%%%%%%%%%%%%%%%%%%%%%%%%%%%%%%%%%%%%%%%%%%%
\subsubsection{Colladon's probable inspiration}
Here is an excerpt from section 3-1 of (Colladon, 1837) entitled Vitesse du son dans les liquides:\\\\
The only experiment carried out until now on the velocity of sound in a liquid body is due to M. Beudant [F.S. Beudant, ~1813]; it was done in sea water, near Marseille, not many years ago. Here are some details of this experiment that this scientist agreed to communicate to us.
	The two observers, separated one from the other by a known distance, had synchronized watches that gave the same time together; at an instant decided beforehand, the person who was supposed to produce the sound raised a flag and simultaneously struck a small immersed bell. The observer situated elsewhere was accompanied by an aid who swam near his boat, heard the sound, and indicated by some sign the moment of its arrival. Thus was obtained the measure of the time taken by the sound to travel from one place to the other: this measurement was not rigorously exact, because the person in the water could not give his signal at precisely the instant that he heard the sound. M. Beaudant concluded from his experiments that the velocity of sound in seawater is 1500 m/s; but as his several experiments led to substantial differences, he  gave this result only as an average.
	It is probable that the real velocity does not differ substantially from this average, which is in fair agreement with theory. But, in order to make this comparison in a more certain manner, it would have been necessary to have a perfectly exact measurement, and, in addition, determine rigorously the density and the compressibility of the liquid at the very temperature of the experiment.

The water of a lake seemed to us to be the most appropriate for giving immediately the velocity of sound in pure water.\\
%%%%%%%%%%%%%%%%%%%%%%%%%%%%%%%%%%%%%%%%%%%%%%%%%%%%%%%%%%%%%%%%%%%%%%%%%%%%%%%%%%%
\subsection{Translation of parts of (Colladon \& Sturm, 1837)}
%%%%%%%%%%%%%%%%%%%%%%%%%%%%%%%%%%%%%%%%%%%ùù
\subsubsection{Contents}
 Introduction\\\\
 Part 1\\
 1-1 Description of the compression instrument\\
 1-2 Compressibility of glass\\
 1-3 Experiments on the compressibility of the liquids\\\\
 Part 2\\
 2-1 Heat liberated by the compession of liquids\\
 2-2 Research on the influence of the compression on the electric conductivity\\\\
 Part 3\\
 3-1 Speed of sound in liquids
 \\\\
 We shall  now translate and comment  parts of  the Introduction, 1-1, 1-3, 2-1 and 3-1.
%%%%%%%%%%%%%%%%%%%%%%%%%%%%%%%%%%%%%%%%%%%%%%%%%%
\subsubsection{Introduction [and Experiments on the compressibility of the liquids]}
This memoir  is divided into four parts.

In the first part, we discuss the experiments relative to the measurement of the compressibility
of fluids; the second part has to do with
experiments on the release of caloric [heat] that accompanies compression; in the third part we seek
to find out if the pressure has an influence on the electric conductivity of these substances;
finally, in the fourth part we give a measure of the velocity of sound in fresh water obtained by
our experiments, and we compare it with the theory.

The first experiments on the compression of fluids were made at the end of the 17th century
by some Florentine physicists. At that time the discoveries of Galileo and Torricelli had attracted
the attention of scientists doing research in experimental physics. Mariotte [(Mariotte, 1717), but also (Boyle, 1662)]
had already recognized the law of the compressibility of gases. The members of the
el Cimento academy,   working together
on a series of experiments concerning the properties of imponderable solids and fluids, rightly
judged that water should be compressible, because it is able to transmit sound, made several tries
to measure this reduction in volume.

In 1761, the physicist John Canton [(Canton, 1761)], returned to this important question.
Having first recognized
the compressibility of water, he undertook precise experiments to measure this compressibility.
His work was not limited to water, since he showed that several other liquids had, like water,
the property of being compressible [(Canton, 1764)].
His experimental method, which since then has been perfected
by Mr. Oersted [(Oersted,1823)], consists in compressing the liquids in instruments resembling thermometers,
composed of a large capacity bulb, surmounted by a capillary tube   opened at its upper
extremity.

Oersted's experiments on water, at constant temperature, did not exceed compressions of
six atmospheres, so that there remained to try larger compressions, not only on water, but
on several other liquids of different densities, and to observe for each of them the influence
of the temperature on the compressibility, as well as to find out if heat is released due to
their compression.

Our instrument for determining the compressibility of liquids is composed of two distinct
parts, one of which  measures the reduction in volume of the liquid which is submitted to a
certain pressure, while the other part quantifies the compression.   The precision of the results
depends on the exact and simultaneous observation of these two quantities.

The method of Canton, perfected by Oersted, is the one we have adopted for our experiments
of compression. It consists, as mentioned previously, in enclosing the liquids in instruments
that we call piezometers, which take the form of large thermometers opened at the top. . . [a
somewhat detailed description of the instrument and the experimental procedure follows].
\\\\
{\it First experiment. Law of the contraction of liquids by increasing compressions}
\\\\
Before entering into the detail of the comparison experiments on different liquids, we
thought it important to determine, by a preliminary experiment carried out with great care,
whether the liquids obey a general law of compression, by which one could predict the results
of experiment, and give a measure--from the condensation observed for a pressure of a small
number of atmospheres--of the condensation that would be produced by an arbitrary pressure.

This research required high precision measurements of the pressure, and especially at
high levels of compression we appealed to the measurement of the rise of the level of mercury
in a barometric tube formed of several parts welded together, and whose overall length was 12.3m.
The lower extremity of this assembly of tubes penetrated into a 0.2m x 0.2m sheet-metal box
filled with mercury. The piston of our compression pump, whose diameter of 27mm and driving
interval of 625mm, was sufficient to raise the mercury to the top of this column whose tubes
were 5mm in diameter. We took care to correct the results of the lowering of the mercury in the
sheet metal box, in relation to the ratio of its diameter to that of the tube. The piezometer
which we used for this experiment, had a  tube which was perfectly cylindrical along a length
of 47cm. The inscribed lines of the graduated scale, divided into half-millimeter steps, were
sufficiently thin for it to be possible to estimate one-fourth millimetre changes.

As the duration of the experiment was rather long, we carried it out at $0^{o}C$ in order to
have a constant temperature during the entire experiment. Here are the results we obtained for
distilled water emptied of air by boiling [table 1 hereafter].

%\newpage
%
\begin{table}[h]
\centering
\begin{tabular}{c|c|c}
Number of atmospheres & half-millimeters on the &
half-millimeters on the\\
~ & scale for compression & scale for decompression
\\\hline
1 & 42 & 42\\\hline
2 & 112 & 115\\\hline
3 & 179 & ~\\\hline
4 & 248 & 250\\\hline
5 & 316 & 319\\\hline
6 & 381 & 384\\\hline
7 & 447 & ~\\\hline
8 & 510 & 714\\\hline
9 & 576 & ~\\\hline
10 & 640 & 645\\\hline
11 & 704 & ~\\\hline
12 & 771 & 774\\\hline
13 & 836 & ~\\\hline
14 & 900 & 902\\\hline
15 & 967 & ~\\\hline
\end{tabular}
\caption{Contraction of distilled water emptied of air (by boiling) at $0^{o}C$}
\label{tvm1}
\end{table}
[The results of table 1 are not exploited numerically, but only serve to show the type of measurements
that were made, and their tendencies, for this and the other liquids].\\\\
Measurement of the contraction of glass [of the capillary tubes of the piezometer].\\\\
Measurement of the contraction of mercury at $0^{o}C$.\\\\
Experiments on distilled water deprived of air by boiling, at $0^{o}C$. Original volume=237,300.
Experiments on [fresh] water not deprived of its air.
\newpage
\begin{table}[h]
\centering
\begin{tabular}{c|c|c|c|c}
Atmosph. & Units on the & Differences & Differences & Contraction for\\
~ & scale  & of pressure & of contraction &  1 atmosphere
\\\hline
1  & 675.50 & ~ & ~ & ~ \\\hline
3  & 653.00 & 2 & 22.50 & 11.250 \underline{}\\\hline
4  & 642.25 & 1 & 10.75 & 10.750 \\\hline
6  & 621.50 & 2 & 20.75 & 10.375 \\\hline
8  & 599.00 & 2 & 22.50 & 11.250 \\\hline
12 & 555.00 & 4 & 44.00 & 11.000 \\\hline
18 & 489.50 & 6 & 65.50 & 10.917 \\\hline
24 & 423.00 & 6 & 66.50 & 11.083 \\\hline
\end{tabular}
\caption{Contractions of [fresh] water not emptied of its air at $0^{o}C$}
\label{tvm1}
\end{table}
This table [table 2] provides us with the same observation as the preceding one [relative to fresh water emptied
of its air], that is that the contractions are constant for equal increases of the pressure. But the
absolute value of the compressibility for one atmosphere is not the same as previously. It is less
than for water deprived of air, so that water containing dissolved air is less compressible than
water deprived of air. We have also verified this result at the temperature of $+4^{o}C$. The ratios of
compressibility were the same. This decrease of compressibility of water that contains dissolved air,
confirms what we already knew, which is that this air is not at all contained in the state of a simple
mixture, but that it is retained by an authentic chemical bond.

The difference of the results obtained by various physicists on the average compressibility of
water, seems to us to be partially due to the fact that they employed water more or less deprived of
air. In effect, a single boiling operation  is not sufficient to eliminate all the air contained in
water; usually this requires three or four such operations.

Before concluding our discussion on this liquid, we should bring to the attention of the reader
that Canton, who measured the compressibility of water not deprived of its air, writes (Trans.Philo.,
1764) that its compressibility was the same as that of water deprived of air; there is no doubt that
the small compressions he employed did not enable the perception of this difference.

These [our] experiments were made with a piezometer for which the weight of a volume of mercury
filling the reservoir was 271,530 mg; the capillary tube was divided into four parts of equal capacity
and the weight of a column of mercury occupying these four parts, was 1578.5 mg. The second part
which was exactly cylindrical, was 344 semi millimeters long.

By comparing the weight of the reservoir with that of the four parts of the capillary tube, one
finds by computation that the volume of the reservoir was equivalent to 233,736 times the equal parts
in capacity of the cylindrical part of the capillary tube corresponding to its length of 344mm.

The liquid, at the beginning of the experiment, filled the reservoir and a portion of the capillary
tube over a length of 680mm. On adding them to the volume of the reservoir that we have just evaluated,
we find that the original volume of the liquid was equal to 237,416 of the small marks on the capillary
tube.

On compressing the liquid, we found that its average contraction was equal to 11 marks of the tube
for each atmosphere, which amounts to 11/237,416  or nearly $46.4\times 10^{-6}$. Such is the contraction observed
for one atmosphere of 0.7466 m of mercury, with the air of the manometer being at the temperature of
$10.25^{o}C$. One must now evaluate the contraction for one atmosphere of 0.76 m of mercury at the
temperature of $10^{o}C$.

As the manometer was maintained at the temperature of $10^{o}C$, each atmosphere is increased by the
increase of temperature of $0.25^{o}C$, whence .25/266 [actually undecipherable], or 1/954. It is then
necessary to diminish its 954th part from the observed contraction, to get the contraction produced
by one atmosphere at $10^{o}C$; this gives 46.35.

The contraction for one atmosphere, at $10^{o}C$, of 0.7466 m of mercury, being
$46.35\times 10^{-6}$, one concludes
that the contraction produced by 1 atm of 0.76 m at $10^{o}C$ is approximately equal to 47.2 parts in a million
of the original volume.

But this is only an apparent contraction. One must add to it, to get the true  contraction, the
volumetric contraction of the glass [of the recipients and capillary tubes] which we evaluated to be 3.3.
One thus obtains 49.5 parts in a million for the true condensation of the liquid, submitted to a pressure
of one atmosphere of 0.76m of mercury at $10^{o}C$.
\newline
\newline
{\it Experiments on alcohol}.\\
{\it  Experiments on sulphuric ether}.\\
{\it  Experiments on water saturated by ammonia}.\\
{\it Experiments on nitric ether at $0^{o}C$}.\\
{\it  Experiments on acetic ether at $0^{o}C$.}\\
{\it Experiments on acetic acid at $0^{o}C$.}\\
{\it Experiments on concentrated sulphuric acid at $0^{o}C$.}\\
{\it  Experiments on nitric acid.}\\
{\it Experiments on turpentine}.
%%%%%%%%%%%%%%%%%%%%%%%%%%%%%%
\subsubsection{Heat released during the compression of liquids}
The phenomena connected with the heat resulting from the compression of substances have, during
the last few years, attracted the attention of several geometers [theoretical physicists and applied
mathematicians] and physicists [experimental and applied physicists]. The understanding of these
phenomena is connected with the most important questions in physics, and could lead to consequences
of great interest concerning the very nature of heat and the relations that exist between this fluid
and the different substances.

These researches have acquired increased importance for the geometers from the moment that Mr.
Laplace demonstrated its application to the theory of sound, and proved that by taking into account
the heat released by the compression of air, one is able to reconcile the mathematical formula
for [the velocity of] sound with the results furnished by experiment.

The phenomena of the release of heat by the compression of gases are now practically elucidated,
thanks to the work of Mr. Gay-Lussac, Mr. Cl\'ement, Mr. Desormes [(Gay-Lussac,1802; Clement \& Desormes, 1819)], and to the recent research of
Mr. de La Rive and Mr. Marcet [(De La Rive \& Marcet, 1827)].

We owe to Mr. Berthollet and Mr. Pictet their observations on the elevation of the temperature
resulting from the compression of metals stamped into medals;  Rumford and Morosi have carried out
research on the heat produced by the friction of metals; but, on account of the extreme difficulty
of such experiments, it is highly improbable that one obtains precise results.

As for the release of heat that should seem to accompany the compression of liquids, it has not
yet been confirmed in direct manner; the only experiments that have been carried out on this subject
are those of Mr. Dessaigne, and the one which was the object of Mr. Oersted's memoir [(Oersted, 1823)]
on the compressibility of water.

The first of these men announced, in a note inserted in the Bulletin de la Soci\'et\'e Philomatique,
that he was able to produce light in several liquids, by submitting them to strong, sudden compressions.
Mr. Oersted says (Annales de Chimie) to have tried without success to produce heat by a compression of
six atmospheres of water. It is doubtful, that he was able, by his experiment,  to accurately measure
the release of heat that should result from the compression of liquids. It would  even be necessary,
to hope to be able to detect it, to employ an instrument able to detect very small amounts of heat,
at the same time,   capable of resisting to considerable
compressions and shocks. The one we have adopted appears to us to satisfy both of these conditions.	
 [Colladon and Sturm then provide the reader with a description of their instrument].

We think to have demonstrated by these experiments: 1) that the temperature of the water does
not sensibly increase by a sudden compression of 40 atmospheres; 2) that for alcohol and sulphuric ether,
the action within a quarter of a second of  compressions of 36 and 40 atmospheres does not raise their
temperature by more than $1^{o}C$; but a much more rapid compression, due to a hammer blow,
releases enough heat to raise the temperature of sulphuric ether by approximately 4 to 6
degrees centigrade.

We shall give, at the end of this memoir, a supplementary proof of the small amount of heat released
by the rapid compression of water, resulting from the comparison of the velocity of sound observed in this
liquid with the formula of Mr. Laplace, independently of any rise in temperature. This comparison will
offer us a precious verification of the experiments contained in this article [Colladon and Sturm mean the
experiments on the release of heat due to compression].

%%%%%%%%%%%%%%%%%%%%%%%%%%%%%%
\subsubsection{Velocity of sound in liquids}
It has been known for a long time that sound propagates through solid and fluid bodies, like air
and air-like fluids. The knowledge of the compressibility of water or of any other liquid, enables
the determination of the velocity of sound therein. Mr. Young [(Young, 1845, p. 287)] and
Mr. Laplace have called attention to an
important application. They have given the formula by which, knowing the degree  of contraction
felt by a liquid in response to an increase of pressure, one can compute the velocity of  sound
propagation in a body of liquid of infinite extent.

The theory, being as complete as it can be, there remained to compare it with experiment, either
to verify one by the other, or to discover a difference that could exist between them. We have
thus undertaken a series of experiments on the velocity of sound in water, the only liquid in
which such experiments are possible, with the aim of comparing the observed velocity with its
theoretical counterpart.

The details of the experimental means and results will be given further on. But before presenting
them, it appears to us opportune to briefly recall the principal points of the theory of sound,
and particularly the formula that serves to compute its velocity in liquid and solid substances.

As we know, Newton is the first to have searched for the laws of the propagation of sound in the
atmosphere [(Newton, 1687; Newton, 1713; Newton, 1726; Newton, 1846)].
He considers an infinitely-long line of molecules in air, and supposes that a portion
of a small subset of this line in air is initially disturbed; he shows that the disturbance
propagates  little by little in all the portions of the air column, as one sees the communication
of movement in a series of elastic balls, and he determines the time that this disturbance, which
produces the sensation of sound, takes to arrive at an arbitrary distance from its origin.  He
finds that the propagation of sound is uniform, and that the velocity of this propagation assumed
to be horizontal, wherein the space that the sound travels in each second, has the value of the
square root of the double product of the height of which gravity makes  bodies to fall in the
first second, by the height of a column of air that would be in equilibrium with a column of
mercury of a barometer, and which would have everywhere the same density as at the bottom of
the column.

Lagrange, Euler, Laplace and M. Poisson [(Lagrange, 1857; Euler, 1955; Laplace, 1816; Laplace, 1822; Laplace, 1823a; Laplace, 1823b; Laplace, 1904; Poisson, 1883)] thereafter deduced the same expression for the
velocity of sound from the analytic partial differential equations that represent the movement
of air, either in an infinitely-long cylindrical column, or in a mass of air of infinite spatial
extent.

By extending their research to the case where the movement of air  takes place in two or three
dimensions, they found that, although the intensity of sound decreases with distance, its velocity
is the same as in the case in which this movement takes place in only one dimension.

However, there existed a notable difference between the velocity of sound in air deduced from this
theory and the one which results from experiments. The physicists, in very great number, who
directly measured this velocity, agreed to have found it to be larger than the computed velocity,
so the much so than the difference attained 1/6-th of the observed value.

We owe to Mr. Laplace [(Laplace,1816), (Laplace,1822), (Laplace,1823a), (Laplace,1823b), (Laplace,1904)],
the real explanation of this difference. It must be attributed to the
increase of elasticity of the air molecules produced by the production of heat which accompanies
their compression. He found that the velocity of sound is equal to the product of the value given
by the formula of Newton, multiplied by the square root of the quotient of the specific heat to
a specific volume. This quotient is a number greater than one. To determine it, M. Laplace
employed the experiments of Mrs. Gay-Lussac [(Gay-Lussac, 1802)] and Welter. The thus-modified formula of Newton
turned out to be more or less in agreement with the real observed velocity.

The computation of the velocity of sound and the laws of its transmission in fluids and solids
are practically the same as in air. It suffices for our purpose to recall here the formula
which represents the velocity of sound in a fluid.

Letting $D$ be the density of a fluid, $K$ the length of a cylindrical  column of
this fluid under known pressure, $\varepsilon$ the small decrease of this length per
given increase of the pressure $P$,  the velocity of sound in this fluid being designated
by $a$ will be given by the following formula
\begin{equation}\label{col1}
a=\sqrt{\frac{PK}{D\varepsilon}}~.
\end{equation}
Let us suppose  that $P$ is the pressure equal to the weight of 76 cm of mercury,
so that $P=0.76g\mathfrak{m}$, $\mathfrak{m}$ designating the density of mercury and $g$ the force of
acceleration of gravity or twice the height that gravity makes  a body to fall in the first second.
The second being taken as the unity of time, we have $g=9.8088$.

The verification of these formulae, as they apply to liquid and solid substances, requires
very precise experiments. The earth does not offer solid masses that are sufficiently
continuous and homogeneous for this type of experiment; it is not probable that one
may verify on a large scale the computations of the velocity of sound in solids.
The experiments of Mr. Biot [(Biot, 1802; Biot, 1816; Roberts, 2008)] on the transmission of sound by iron pipes taught that its velocity
is much greater than that of its transmission through air; but, as the sound arrived in less
than a half of a second, one could only deduce from this a very uncertain evaluation, which
could not be looked-at as sufficient to verify the formula.

Water appears to us to be the only substance in which such experiments can be carried
out precisely: it is accepted that this liquid transmits sound over very great distances.
Franklin discovered that the sound of two stones hitting each other under water can be
heard a half mile away. However, it does not appear that he thought of measuring the velocity.

With this aim, one of us (Mr. Colladon) went to Switzerland in the month of October 1826,
to undertake, in the lake of Geneva, a series of experiments on the propagation of sound in
pure water, and to determine the velocity of this transmission, which had not been measured
until then.

We first did some trials (As Mr. Sturm was unable to accompany me to Geneva to help me in
these experiments, I was obliged to tell the story of this research on the propagation of
sound in my sole name (D.C.). To determine the best manner of producing  sounds in water
that could be heard at great distances we first tried the explosion of  gunpowder, then
a violent shock on an immersed anvil, and finally  hammer blows on a bell suspended in the
water: this latter means was judged to be the best one. Each blow on the bell produced a very
brief sound with an easy-to-distinguish  metallic timbre.

However, when at this same distance, one plunged his head entirely in the water, he heard
very distinctly each blow; by increasing the distance, the sound conserved enough intensity
to be heard until 2009 m. It was in this manner that the first attempts of obtaining a measure
of the velocity of sound were made (I made my first experiments with the help of  Mr. A. de
Candolle, near the countryside of his father, situated on the rim of the lake). These
experiments were carried out during the night and were very painstaking. The person who
listened,  not being able to perceive the signals, communicated to a second observer the
announcement of a sound; this second person noted on a watch the interval of time between
the perception of the signals designed to fix the instant of the blow and the arrival of
the sound. This procedure was not very precise; the intermediary whose role was to hear
the sound could not announce its arrival sufficiently promptly for this not to result in
some error. These errors were compounded by the fact that the greatest  distance at which
the blows to the bell could be distinguished was only 2500 m, and this distance was traveled
by the sound in less than two seconds.

These difficulties suggested to me the idea of searching for a different means of listening
to sounds in water; several tries enabled me to discover an instrument that I think is new,
and which helped me to repeat the experiments at a distance of 14000 m.

I now briefly explain the principle on which is based its construction.

We said that sound waves transmitted by the liquid do not communicate with air in the
 neighborhood of the bell; when the direction of these waves meets the surface at a very
 acute angle, they reflect towards the interior of the liquid mass without communicating any
 noticeable disturbance to the air in contact with this surface.	It appeared to me probable
 that if one could break the continuity of this mass by introducing therein a metallic thin-walled
 vase filled with air, the gas contained in this envelope could receive and transmit to the exterior
 the vibratory motion propagated in the liquid.

In my first trials, I employed a simple tin plate tube with prismatic cross-section; this
tube was about three meters in length and 15cm on each side; it was closed at its bottom
extremity, and to  the lower end was attached a ring to which was suspended a weight sufficient
to keep the instrument immersed at a depth of two meters; the upper extremity was open and
situated at a height of  one meter above the water. During the first experiment with this
instrument I was distant of more than two thousand meters from the bell; when the blows
were made, one heard very distinctly the sound exit the tube, and he  could have thought
that the sound came from the blow of a small metallic body against the bottom of the tube;
and it was sufficiently strong for it to be heard at a distance, and at more than two meters
from the opening he still could make out every blow.

I undertook to improve this instrument, and adopted for my last trials a long cylindrical
tin-plate tube, curved on its upper end and terminated by a small opening  at which one
applied the ear. The lower part of the tube was likewise curved, with its extremity
entirely closed by a tin-plate disk [in the vertical position, i.e., facing the sound source;
note that if the vertically-oriented tube was not curved, its opening (and a tin-plate disk to
close it) at its bottom end would be horizontal].

This instrument   increases to such an extent the sound sensation, that the noise  produced
by a blow to a bell heard in this device at 14000 m  appeared to me as intense as  the same
sound heard at 200m by simply immersing one's head.

It is very probable that by increasing considerably the dimensions of the device the latter
could serve for underwater communication over considerable distances (in the instrument that
I used, the cross section was approximately twenty square decimeters and the tube length five
meters; the bell weighed 65 kg). I am convinced that by employing a heavier bell and by
improvement and increase of the size of the listening device, one could succeed in communicating
easily, under the water of a lake or of sea, at a distance of fifteen  or twenty lieues [60-80km].

I must insist on the fact that one would hear absolutely nothing if the instrument were not
 closed and entirely filled with air; I assured myself of this through several experiments.

One could have thought that the sound  be heard in the helmet of a professional diver;
 however, I made some experiments in Rouen, in 1830 which gave rise to no satisfactory result
 since  the blows against a bell of the same size as the one I used in the Geneva lake could
 not be heard at a distance of two or  three hundred meters. It is probable that this remarkable
 effect should be attributed to the thickness of the helmet walls: those of our diver helmet
 were in cast iron and were approximately twelve cm thick.

Having recognized the possibility of hearing a sound at several lieues, I undertook a new
 series of experiments on the velocity of sound, between two locations situated in the small
 towns of Rolle and Thonon, at opposite banks of the lake, in a spot where it is the largest.
 The distance between these two towns is approximately fourteen thousand meters (this distance
 is approximately half that between Montlh\'ery and Montmartre, chosen by the members of the
 French Academy in 1738 to measure the velocity of sound in air).

This position is very favorable for the measurements; the distance included between the two
 towns can be exactly verified by attaching it to that of Geneva to Langin, which served as the
 base for the triangulation of the L\'eman valley; the average depth of the water is very large
 between the two banks. The bottom meets the banks on both sides at substantially the same angle,
 and no intermediate shallow bottom exists that could intercept the sound.  The average depth of
 the lake between Rolle and Thonon is one hundred fourty meters;  moreover, one finds in this
 interval no trace of current; the water is  remarkably transparent and not bothered by the
 agitation of waves.

We were obliged to make some modifications to the means employed previously to indicate the
 instant of the blow to the bell.

The curvature of the earth between the two stations is such that from one of these points,
  one could not see objects placed on the other bank near the water surface. One can surmount
  this difficulty by using gunpowder signals; the flame of this gunpowder was not seen
  [recall that the experiments were carried out at night] from
  the other station, but this sudden light produced a distinct luminous burst that appeared
  to rise by several degrees above the horizon each time that the quantity of burnt powder
  exceeded one hundred fifty grams. [A very simple mechanism was contrived to make the]
  gunpowder take fire at the precise instant the hammer reaches  the bell.

The experiments carried out by this process have acquired such a regularity that in the
   four or five last series of measurements, the largest difference never exceeded a half
   second (in all the experiments done between Rolle and Thonon, we always proceeded as
   follows with the measurements; we had at the first station a watch which was synchronized
   with, and gave the same time as the watch at the other station, and the experiments were
   done during fifteen minutes, every fifteen minutes. To avoid the possibility that the sound of the
   bell  be confounded with foreign noise, we always struck three blows with an interval
   of one second; the last two served uniquely to verify the nature of the sound, and were not
   accompanied by a luminous signal. I suppressed, in these series, all the experiments in which
   the initial blow was not followed by the sound of the two additional blows.

I was seated [in a boat] at the other station, my face turned towards the [location of the
   other boat from which was suspended the] bell and my head pressed against the opening  of
   the [tin plate] tube, which a helper maintained in this position; I thus could use my two
   hands to hold and stop the stopwatch, and I could easily observe the gunpowder signals and
   hear the arrival of the sound [from the struck bell].

The stopwatch which I used was accurate to a quarter of a second and had a sensitive trigger;
at the moment  the gunpowder took fire I pressed the trigger to start the needle movement, and
stopped it at the arrival of the sound. The angular interval covered on the dial indicated the
time taken by the sound to arrive.

A small interval of time necessarily  occurred between the moment of seeing the flash of light
and the moment of pressing the trigger. A similar delay occurred in the perception of the sound;
but this second delay was probably a little bit smaller [because the instant of  sound was expected,
whereas the instant of the luminous signal was relatively unexpected].

We found [in the series of experiments] that the duration between the perception of the light
flash and the arrival of the sound is larger than 9s and smaller that 9,5s; its average value
is a little bit above 9,25s. If we estimate to less than a quarter second  the small error
mentioned above, we can adopt 9,4s as the time really taken by the sound to go from one station
to the other.

[Triangulation gave] 13887 meters for the distance between the two opposing banks. 	By subtracting
400 meters for the distances of the  boats from the two shores, one gets 13487 meters for the
distance between the two stations. This number can be regarded as exact to less than 20 meters.

By dividing the distance of 13487 meters by the time 9.4 s, one obtains the velocity of sound
in [Lake Geneva] water, to be $1435\pm 24~ms^{-1}$ [for an average temperature of the body of
water, at the depth of emitter and receiver, of $8.1^{o}C$].

To compare these results with those of the computation, it was necessary to determine with much
care the compressibility of this water at that temperature, as well as the quotient of its density
to that of distilled water at $0^{o}C$.

The water of the lake, at a distance sufficiently far from the inlet of the Rhone river, can
be considered to be perfectly pure; it hardly contains more than 1/2000 of its weight in
foreign matter.

The density of the lake water is approximately 1.00015 at $4^{o}C$, that of distilled water
at this temperature being taken as unity. And as the volume of water increases by 0.00013 when
it goes from $4^{o}C$ to $8^{o}C$, the density of the water in which was measured the velocity
of sound was one, plus a negligible fraction.

However small was the quantity of foreign matter contained in this water, we felt it necessary
to determine directly its compressibility, instead of supposing it to be equal to that of
distilled water. We made this measurement on water taken at the surface at a location in
between the two stations; this water was introduced into a high-quality piezometer, with
the precautions indicated previously concerning the compression of water saturated with air.

Let us now return to the formula for the speed of sound given above, in order to substitute
therein the values we have just determined; this formula is
\begin{equation}\label{2}
 a=\sqrt{\frac{P}{D\varepsilon/K}}~.
\end{equation}
The (measured) quantities designated by $D,\varepsilon/K,P$, for the water of the Lake of
Geneva at the temperature of $8.1^{o}C$, are:
\begin{equation}\label{3}
D=1~~~;~~~\varepsilon/K=48.66\times 10^{-6}~.
\end{equation}
[Note that as a result of his laboratory experiments on fresh water not deprived of air
(see table 2 above and the associated comments), Colladon found $\varepsilon/K=49.5\times 10^{-6}$; the
difference of this value from $48.66\times 10^{-6}$ is probably due to the fact that the latter figure
is for lake Geneva water at $8.1^{o}C$ whereas the former is for distilled water at $0^{o}C$].

If we take for $P$ the pressure of one atmosphere of 0.76m of mercury at the temperature
of $10^{o}C$, at which our manometer was fixed,
designating by $\mathfrak{m}$ the density of this mercury and by $g=9.8088ms^{-2}$ the
accelerating force of gravity, or the double of the height which it makes bodies to fall
in the first second, we have
\begin{equation}\label{4}
P=(0.76)*g\mathfrak{m}~.
\end{equation}
The density of mercury at $0^{o}C$ is, by virtue of the experiments of Mr. Dulong and Mr.
Petit [(Dulong \& Petit, 1816)], equal to 13.568; that of distilled water at $3.90^{o}C$
being taken equal to unity.
Also, the dilatation of mercury is 0.00018 for each increase of one degree of temperature,
so that it is 0.0018 for $10^{o}C$. Thus, mercury, in going from $0^{o}C$ to $10^{o}C$, increases
in volume from 1 to 1.0018.

The density of mercury at $10^{o}C$ will then be equal to its density at $0^{o}C$ or 13.568 divided
by 1.0018; so that we have
\begin{equation}\label{5}
\mathfrak{m}=\frac{13.568}{1.0018}=13.544~.
\end{equation}
Thus, substituting all these values in the formula for the velocity of sound, gives:
\begin{multline}\label{6}
D=1 kg/cm^{3} =10^{3}kg/m^{3}~~;~~\varepsilon/K=48.66\times 10^{-6}~~;~~\\
P=(0.76)(9.8088)(13.544)=100.9663kg~m^{-1}s^{-2}~.
\end{multline}
or
\begin{equation}\label{7}
a=1437.8~ms^{-1}~.
\end{equation}
[Actually, this result is in error, and should be $a=1440.5~ms^{-1}$;
if we adopt the formula, in modern notation, $a=\sqrt{1/\rho\kappa_{S}}$,
with density $\rho=1~kg~m^{-3}$, in order to get  $a=1440.5ms^{-1}$,
the adiabatic compressibility must be $\kappa_{S}=0.48192\times 10^{-9} ms^{2}kg^{-1}=
0.48192~GPa^{-1}$, which is consistent with the presently-known
fact (Chaplin, 2012) that (presumably for fresh water at 1 atm pressure) at $0^{o}C$
the adiabatic compressibility is  $0.5086~GPa^{-1}$ at $0^{o}C$ and
$0.4776~GPa^{-1}$ at $10^{o}C$].

This is the theoretical determination of the velocity of sound in water deduced
from the density and the compressibility of this liquid, under the hypothesis that
no heat is produced by the rapid compression of the liquid molecules that could
increase their temperature.

In our experiments the distance 13487m was traveled in 9.4s, which gives for the measured velocity
\begin{equation}\label{7}
 a=13487/9.4=1435~ms^{-1}~,
\end{equation}
[actually, $1434.8~ms^{-1}$], so that the two values differ by no more than three $ms^{-1}$.
This remarkable coincidence, by confirming the observations contained in the second part of
this memoir, can serve to show better than any direct experiment that the compression of water
does not make its temperature vary [this is consistent with the fact (Chaplin, 2012) that the
isothermal and adiabatic compressibilities ($\kappa_{T}$ and $\kappa_{S}$ respectively) of fresh water at 1 atm
are very close both at $0^{o}C$ in which case they are $0.5089~GPa^{-1}$ and
$0.5086~GPa^{-1}$ respectively, as well as at $10^{o}C$ in which case they are
$0.4781~GPa^{-1}$ and $0.4776~GPa^{-1}$ respectively].
%%%%%%%%%%%%%%%%%%%%%%%%%%%%%%%%%%%%%%%%%%%%%%%%%%%%%%%%%%%%%%%%%%%%%%%%%%%%%%%%%%%%%%%%%%%%%%%%%%%%%%%%%%%%%%%%%%%%%%%%%%%%%%%%%%%%%%%%%%%%%%%%%
%%%%%%%%%%%%%%%%%%%%%%%%%%%%%%%%%%%%%%%%%%%%%%%%%%%%%%%%%%%%%%%%%%%%%%%%%%%%%%%%%%%%%%%%%%%%%%%%%%%%%%%%%%%%%%%%%%%%%%%%%%%%%%%%%%%%%%%%%%%%%%%%%
\section{An inverse  problem of wave-like nature}
The retrieval of the velocity of sound   is an inverse wave propagation problem which can be formulated as follows:\\
i) trace a line from the location of a pulse-like sound-emitting source $P_{s} $ to the location $P_{r}$ of a hypothetical, rather distant (from the source location) receiver,\\
ii) place two identical receivers, one at a point $P_{1}$ not too distant laterally from  the line and longitudinally from $P_{s}$, and the other at a point $P_{2}$ not too distant laterally from the line and longitudinally from $P_{r}$,\\
 iii) observe the coordinates of the two points $P_{1}$ and $P_{2}$,\\
  iv)  register the sound signals at these two points and note the time instants at which  the same remarkable feature, such as the maximum, of the pulse occurs,\\
  v) either by comparison of the registered pulses at the two locations, or from the measured time interval between the occurrence of the remarkable feature in the two signals, retrieve, using the coordinate information of the receiver locations, the velocity $v$ of the medium in which the sound propagates, under the assumptions that this medium is a homogeneous fluid of infinite spatial extent, the support of the source is line-like, and the spectrum and onset time of the sound pulse is known (note that in Colladon's experiments in Lac Leman, these assumptions were made, but no account was taken of a possible lack of information concerning the pulse spectrum and of the onset time of the pulse).
%%%%%%%%%%%%%%%%%%%%%%%%%%%%%%%%%%%%%%%%%%%%%%%%%%%%%%%%%%%%
\subsection{Wave equation}
Consider infinite space $\mathbb{R}^{3}$ wherein a point $P$ is defined by the cartesian coordinates $(x,y,z)$  relative to the origin $O$ at $(0,0,0)$. This (unbounded) space is occupied by a homogeneous, inviscid fluid in which is present a source of density $S$ giving rise to a radiated pressure field $U$. Let us assume that $S$: i) is of bounded support, ii) is localized near $O$, and iii) depends only on $x,y$ as well as  on the time variable $t$. It ensues that the scalar field $U$ depends only on $\mathbf{x}:=(x,y)$ and $t$.
$U(\mathbf{x},t)$ obeys:
\begin{itemize}
\item the wave equation
\begin{equation}\label{pp1}
    U_{,tt}(\mathbf{x},t)-v^{2}[U_{,xx}(\mathbf{x},t)+U_{,yy}(\mathbf{x},t)]=-S(\mathbf{x},t)
    ~;~\mathbf{x}\in\mathbb{R}^{2}~,~t\in\mathbb{R}^{+}~,
\end{equation}
(wherein $F_{,\xi\xi}:=\frac{\partial^{2} F}{\partial\xi^{2}}$ and $v$ is the bulk phase velocity in the solid (fluid)) and
 \item the radiation condition
\begin{equation}\label{pp2}
    U(\mathbf{x},t)\rightarrow 0~;~(r=\sqrt{x^{2}+y^{2}},~t)\rightarrow\infty ~.
\end{equation}
\end{itemize}

Note that:\\
1- due to the homogeneous, inviscid nature of the fluid, $v$ is a {\it real constant},\\
2- both $S$ and $U$ are {\it real} functions.\\

We expand $U$ and $S$ in  Fourier integrals
\begin{equation}\label{pp3}
    S(\mathbf{x},t)=\int_{-\infty}^{\infty}s(\mathbf{x},\omega)\exp(-i\omega t)d\omega~,~U(\mathbf{x},t)=\int_{-\infty}^{\infty}u(\mathbf{x},\omega)\exp(-i\omega t)d\omega~,
\end{equation}
(wherein $\omega$ is a angular frequency) which, introduced into (\ref{pp1}), give rise to the inhomogeneous Helmholtz equation
\begin{equation}\label{pp4}
    u_{,xx}(\mathbf{x},\omega)+u_{,yy}(\mathbf{x},\omega)+k^{2}u(\mathbf{x},\omega)=-s(\mathbf{x},\omega)~;~\mathbf{x}\in\mathbb{R}^{2}~,
\end{equation}
wherein
\begin{equation}\label{pp5}
    k:=\frac{\omega}{v}~,
\end{equation}
$k$ and $v$ being the wavenumber and velocity (both being real and independent of $\omega$) in the medium.

The previously-underlined real nature of $S$ and $U$, imposes the following constraints on the functions $s$ and $u$:
\begin{equation}\label{pp5a}
   s(\mathbf{x},-\omega)=s^{*}(\mathbf{x},\omega)~~,~~u(\mathbf{x},-\omega)=u^{*}(\mathbf{x},\omega)
~,
\end{equation}
wherein $^{*}$ denotes the complex conjugate operator. It then ensues
\begin{equation}\label{pp5c}
   S(\mathbf{x},t)=2\Re\int_{0}^{\infty}s(\mathbf{x},\omega)\exp(-i\omega t)d\omega~~,~~ U(\mathbf{x},t)=2\Re\int_{0}^{\infty}u(\mathbf{x},\omega)\exp(-i\omega t)d\omega~.
\end{equation}
%
%%%%%%%%%%%%%%%%%%%%%%%%%%%%%%%%%%%%%%%%%%%%%%%%%%%%%%%%%%%%%%%%%%%%%%%%%%%%
\subsection{The space-frequency  expression of the 2D free-space Green's function}\label{sfgf}
Consider the case of a line source located at $\mathbf{x}'=(x',y')$
\begin{equation}\label{pp6}
   s(\mathbf{x},\omega)=\delta(x-x')\delta(y-y')~,
\end{equation}
wherein $\delta(.)$ is the Dirac delta distribution. The  possible solutions of (\ref{pp4}) are (Morse \& Feshach, 1953)
\begin{equation}\label{pp7}
   u(\mathbf{x},\omega):=u^{(1)}(\mathbf{x},\omega)=\frac{i}{4}H_{0}^{(1)}(k|R|)~,
\end{equation}
and
\begin{equation}\label{pp7a}
   u(\mathbf{x},\omega):=u^{(2)}(\mathbf{x},\omega)=-\frac{i}{4}H_{0}^{(2)}(k|R|)~,
\end{equation}
wherein  $H_{0}^{(j)}$ is the zeroth-order Hankel function of the $j$-th kind, and
\begin{equation}\label{pp8}
   R=\sqrt{(x-x')^{2}+(y-y')^{2}}~.
\end{equation}

We make use of the asymptotic forms (Abramowitz \& Stegun, 1968) of the Hankel functions
to find the two asymptotic solutions (i.e., for $j=1,2$)
\begin{equation}\label{pp10}
   u^{(j)}(\mathbf{x},\omega)\sim \frac{(3-2j)i}{4}\sqrt{\frac{2}{\pi k|R|}}\exp\left[(3-2j)\left(ik|R|-\frac{i\pi}{4}\right)\right]~;~k|R|\rightarrow\infty~.
\end{equation}

The $j=1$ ($j=2$) solution  represents an {\it outgoing wave}, which in the space-time domain satisfies the radiation condition (\ref{pp2}) from the line source located at $\mathbf{x}'$ for $\omega\geq 0$ ($\omega\leq 0$ respectively), so that the field radiated by this source, satisfying the radiation condition, is (the so-called space-frequency 2D free space Green's function of the Helmholtz operator)
\begin{equation}\label{pp11a}
   u(\mathbf{x},\omega):=G^{(1)}(\mathbf{x},\mathbf{x}';\omega)=
   \frac{i}{4}H_{0}^{(1)}\left(\frac{\omega}{v}\|\mathbf{x}-\mathbf{x}'\|\right)~;~\forall \omega\geq 0~,
\end{equation}
\begin{equation}\label{pp11b}
   u(\mathbf{x},\omega):=G^{(2)}(\mathbf{x},\mathbf{x}';\omega)=
   \frac{-i}{4}H_{0}^{(2)}\left(\frac{\omega}{v}\|\mathbf{x}-\mathbf{x}'\|\right)~;~\forall \omega\leq 0~.
\end{equation}
A noteworthy property of this Green's function, consistent with (\ref{pp5a}), is (due to the facts (Abramowitz \& Stegun, 1968)  that $H_{0}^{(1)}(-z)=H_{0}^{(1)}(z)$ and $[H_{0}^{(1)}(z)]^{*}=H_{0}^{(2)}(z)$):
\begin{equation}\label{pp11c}
   G^{(j)}(\mathbf{x},\mathbf{x}';-\omega)=[G^{(j)}(\mathbf{x},\mathbf{x}';\omega)]^{*}~.
\end{equation}
%
%%%%%%%%%%%%%%%%%%%%%%%%%%%%%%%%%%%%%%%%%%%%%%%%%%%%%%%%%%%%%%%%%%%%%%%%%%%%
\subsection{The space-frequency  expression of the field radiated
by general 2D sources}\label{sftds}
%
%%%%%%%%%%%%%%%%%%%%%%%%%%%%%%%%%%%%%%%%%%%%%%%%
\subsubsection{The case of arbitrary-support 2D sources}
Suppose the support of the source is a finite subset of $\mathbb{R}^{2}$ localized at $\mathbf{x}_{0}=(x_{0},y_{0})$.

The solution of the inhomogeneous Helmholtz equation, for a more general (than (\ref{pp6})) 2D source distribution, satisfying the radiation condition,  takes the form
\begin{equation}\label{pp13}
  u(\mathbf{x},\omega)=\int_{\mathbb{R}^{2}}G^{^{(1)}}(\mathbf{x},\mathbf{x}';\omega)s(\mathbf{x}',\omega)dx'dy'~;
  ~\forall\omega\geq 0~.
\end{equation}

From here on, we shall consider sources that can be cast in the form
\begin{equation}\label{pp14}
  s(\mathbf{x},\omega)=\mathfrak{h}(\omega)\mathfrak{g}(\mathbf{x})~,
\end{equation}
wherein $\mathfrak{h}(\omega)$ is the spectrum of the space-frequency representation of the source. In order to satisfy the constraint (\ref{pp5a}), we must have
\begin{equation}\label{pp15}
  \mathfrak{h}(-\omega)=\mathfrak{h}^{*}(\omega)~.
\end{equation}
From here on (in order to not unduly lengthen our paper), we shall only consider sources of vanishing geometrical support, although, in order for the theoretical analysis of the Colladon experiment to be complete, we should have taken into account the non-zero nature of this support.
%%%%%%%%%%%%%%%%%%%%%%%%%%%%%%%%%%%%%%%%%%%%%%%%%%%%%%%%%%%%%%%%%%%%%%%%%%%
\subsubsection{The space-frequency domain expression of the field radiated
by a line source in the far-field zone}\label{fgf}
The (line) source, located at $\mathbf{x}_{0}=(x_{0},y_{0})$, is characterized by
\begin{equation}\label{fgf0}
   s(\mathbf{x},\omega)=\mathfrak{h}(\omega)\delta(x-x_{0})\delta(y-y_{0})~,
\end{equation}
The region of $\mathbb{R}^{2}$ corresponding to $k|\mathbf{x}-\mathbf{x}_{0}\|>>1$ is called the {\it far-field zone}.
The exact expression of the field in this zone (and elsewhere) is
\begin{equation}\label{fgf1}
 u(\mathbf{x},\omega)=\int_{\mathbb{R}^{2}}G^{^{(1)}}(\mathbf{x},\mathbf{x}';\omega)s(\mathbf{x}',\omega)dx'dy'=
 \mathfrak{h}(\omega)\frac{i}{4}H_{0}^{(1)}\left(k\|\mathbf{x}-\mathbf{x}_{0}\|\right)~;
  ~\forall\omega\geq 0~.
\end{equation}
We make use of (\ref{pp10}) to obtain
\begin{equation}\label{fgf2}
   u(\mathbf{x},\omega)\approx\mathfrak{h}(\omega)\frac{i}{4}\sqrt{\frac{2}{\pi k\|\mathbf{x}-\mathbf{x}_{0}\|}}\exp\left[\left(ik\|\mathbf{x}-\mathbf{x}_{0}\|-\frac{i\pi}{4}\right)\right]~;
   ~k|\mathbf{x}-\mathbf{x}_{0}\|>>1~.
\end{equation}
%
%%%%%%%%%%%%%%%%%%%%%%%%%%%%%%%%%%%%%%%%%%%%%%%%%%%%%%%%%%%%%%%%%%%%%%%%%%%
\subsubsection{The space-frequency  expression of the field radiated
by a line source in the paraxial region of the far-field zone}\label{fgfp}
Assuming that
\begin{equation}\label{fgfp1}
|y-y_{0}|<<|x-x_{0}|~,
\end{equation}
we obtain the approximation
\begin{equation}\label{fgfp2}
\|\mathbf{x}-\mathbf{x}_{0}\|\approx |x-x_{0}|
\left[1+\frac{1}{2}\left(\frac{y-y_{0}}{x-x_{0}}\right)^{2}\right]~,
\end{equation}
so that
\begin{equation}\label{fgfp3}
u(\mathbf{x},\omega)\approx\mathfrak{h}(\omega)\frac{i}{4}\sqrt{\frac{2}{\pi k|x-x_{0}|}}\exp\left\{ik\left[|x-x_{0}|+\frac{1}{2}\frac{(y-y_{0})^{2}}{|x-x_{0}|}\right]
-\frac{i\pi}{4}\right\}
~,
\end{equation}
which is the approximation of the space-frequency field produced by a line source in the paraxial region of the far-field zone.
%\newpage
%%%%%%%%%%%%%%%%%%%%%%%%%%%%%%%%%%%%%%%%%%%%%%%%%%%%%%%%%%%%%%%%%%%%%%%%%%%
\subsubsection{Numerical comparison of the exact, far-field and paraxial far-field space-frequency   fields radiated
by a line source}\label{ncspf}
The following two  figures apply to the case: $x_{0}=y_{0}=0$, $y=0$, $v=1500~m/s$. $\omega$ is in units of Hz.

\begin{figure}
[ht]
\begin{center}
\includegraphics[scale=0.4]{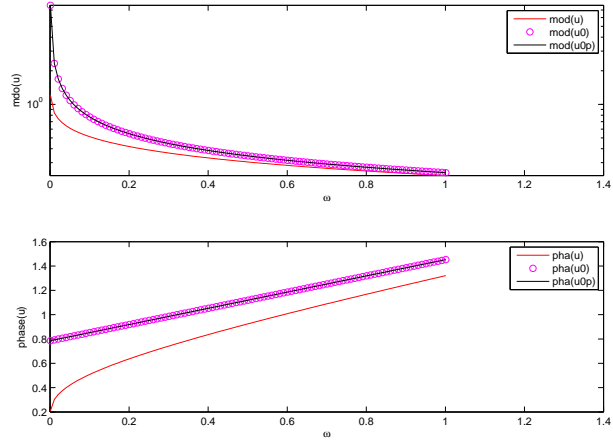}
\caption{Modulus (top panel) and phase (bottom panel) of the exact (U), far-field approximation (U0p) and paraxial far-field approximation (U0) space-frequency fields at $x=1000m$  as a function of $\omega$ (in units of Hz) for low frequencies.} \label{ncspf-3}
\end{center}
\end{figure}

\begin{figure}
[ht]
\begin{center}
\includegraphics[scale=0.4]{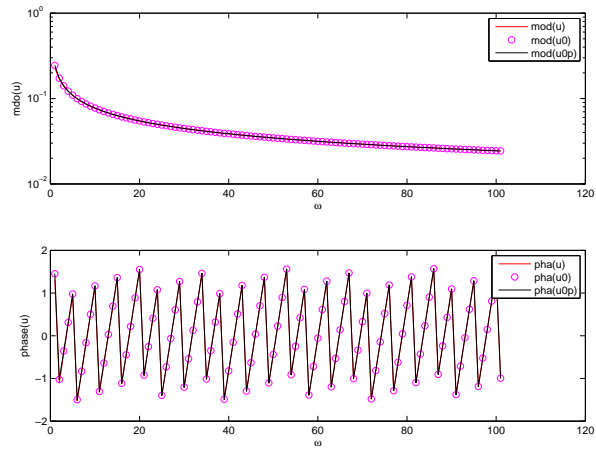}
\caption{Modulus (top panel) and phase (bottom panel) of the exact (U), far-field approximation (U0p) and paraxial far-field approximation (U0) space-frequency fields at $x=1000m$   as a function of $\omega$ (in units of Hz) for high frequencies.} \label{ncspf-6}
\end{center}
\end{figure}

Figs. (\ref{ncspf-3}) (low frequencies) and (\ref{ncspf-6}) (high frequencies), relative to the source-to-observer distance $1000m$, depict the numerical predictions of the space-frequency fields obtained by the rigorous theory, the far-field approximation, and the paraxial far-field approximation. As expected, the paraxial far-field approximation is appropriate only for large frequencies (all the larger the smaller the distance between the source and receiver).
\clearpage
\newpage
%%%%%%%%%%%%%%%%%%%%%%%%%%%%%%%%%%%%%%%%%%%%%%%%%%%%%%%%%%%%%%%%%%%%%%%%%%%%%%%
\subsection{The space-time expression of the 2D free-space Green's function}\label{gf}
\begin{figure}
[ht]
\begin{center}
\includegraphics[scale=0.4]{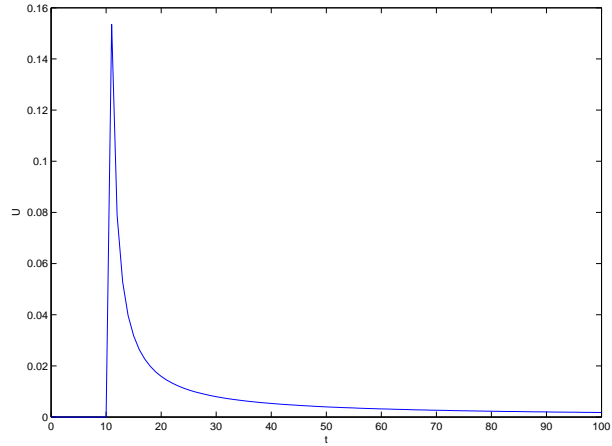}\hfill
\caption{Time history of pressure $U$ at a given $\mathbf{x}=(300~m,300~m)$. Abscissa is  $t$ in sec.} \label{gf-1}
\end{center}
\end{figure}

\begin{figure}
[ht]
\begin{center}
\includegraphics[scale=0.4]{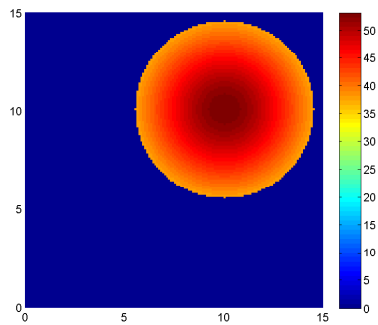}\hfill
\caption{Map (vertical axis is $y$ and the horizontal axis is $x$) of pressure $U$ at a given $t=10.003~s$.} \label{gf-2}
\end{center}
\end{figure}

The (line) source of the Green's function is characterized, in the space-time domain, by
\begin{equation}\label{gf1}
  S(\mathbf{x},t)=\delta(x-x_{0})\delta(y-y_{0})\delta(t-t_{0})~,
\end{equation}
wherein $\delta(~)$ is the Dirac delta distribution,  $\mathbf{x}_{0}=(x_{0},y_{0})$  the location of the line source, and $t_{0}$ the instant at which it is turned on. The space-frequency representation of this source is, on account of the sifting property of the Dirac delta distribution,
\begin{equation}\label{gf3}
  s(\mathbf{x},\omega)=\frac{\exp(i\omega t_{0})}{2\pi}\delta(x-x_{0})\delta(y-y_{0})~,
\end{equation}
whence
\begin{equation}\label{gf4}
  \mathfrak{h}(\omega)=\frac{\exp(i\omega t_{0})}{2\pi}~,
\end{equation}
is the spectrum of the source. This spectrum has two remarkable properties: it is complex {\it constant for all frequencies}, and satisfies the constraint (\ref{pp14}). It follows that:
\begin{equation}\label{gf5}
  U(\mathbf{x},t)=2\Re\left[\frac{i}{8\pi}\int_{0}^{\infty}
 H_{0}^{(1)}\left(\frac{\omega}{v}\|\mathbf{x}-\mathbf{x}_{0}\|\right)e^{-i(\omega-\omega_{0})t}d\omega\right]~,
\end{equation}
or, making use of the fact (Abramowitz \& Stegun, 1986) that $H_{0}^{(1)}(z)=J_{0}(z)+iY_{0}(z)$, with $J_{0},~Y_{0}$ the zeroth-order Bessel and Neumann functions respectively,
\begin{equation}\label{gf6}
  U(\mathbf{x},t)=\mathcal{J}(\mathbf{x},t)+\mathcal{Y}(\mathbf{x},t)~,
\end{equation}
wherein
\begin{multline}\label{gf7}
  \mathcal{J}(\mathbf{x},t)=\Re \frac{i}{4\pi}\int_{0}^{\infty}
 J_{0}\left(\frac{\omega}{v}\|\mathbf{x}-\mathbf{x}_{0}\|\right)e^{-i(\omega-\omega_{0})t}d\omega~~,\\
  \mathcal{Y}(\mathbf{x},t)=-\Re\frac{1}{4\pi}\int_{0}^{\infty}
 Y_{0}\left(\frac{\omega}{v}\|\mathbf{x}-\mathbf{x}_{0}\|\right)e^{-i(\omega-\omega_{0})t}d\omega~.
\end{multline}
We make use of the following  results in (Abramowitz \& Stegun, 1986):
\begin{equation}\label{gf12}
  \int_{0}^{\infty}J_{0}(a\omega)e^{-ib\omega}d\omega=\left\{
\begin{array}{cc}
\frac{1}{\sqrt{a^{2}-b^{2}}} & ;~~~0\leq b<a\\
\frac{-i}{\sqrt{b^{2}-a^{2}}}~ & ;~~0<a<b
\end{array}
\right.
 ~,
\end{equation}
\begin{equation}\label{gf13}
  \int_{0}^{\infty}Y_{0}(a\omega)e^{-ib\omega}d\omega=
  \left\{
\begin{array}{cc}
\frac{-2ia}{\pi\sqrt{a^{2}-b^{2}}}\sin\left(\frac{b}{a}\right) & ;0\leq b<a\\
\frac{-1}{\sqrt{b^{2}-a^{2}}}+\frac{-2i}{\pi\sqrt{b^{2}-a^{2}}}\ln\left[\frac{b-\sqrt{b^{2}-a^{2}}}{a}\right]^{*} & ;0<a<b
\end{array}
\right.
 ~,
\end{equation}
 to obtain the final result
\begin{equation}\label{gf16}
 U(\mathbf{x},t)=
  \left\{
\begin{array}{cc}
0 & ;~~0\leq t-t_{0}<\frac{\|\mathbf{x}-\mathbf{x}_{0}\|}{v}\\
\frac{1}{2\pi}\frac{1}{\sqrt{(t-t_{0})^{2}-\frac{\|\mathbf{x}-\mathbf{x}_{0}\|^{2}}{v^{2}}}} & ;~0<\frac{\|\mathbf{x}-\mathbf{x}_{0}\|}{v}<t-t_{0}
\end{array}
\right.
 ~,
\end{equation}
Two graphical representations of this function (which vanishes for infinite time and distance from the source as it should by virtue of the radiation condition (\ref{pp2})) are given in figs. (\ref{gf-1}) and (\ref{gf-2}) in which $x_{0}=y_{0}=10~m$, $v=1500~m/s$ and $t$ is in units of seconds.
\clearpage
\newpage
%%%%%%%%%%%%%%%%%%%%%%%%%%%%%%%%%%%%%%%%
\subsubsection{A method for retrieving $v$ from the space-time Green's function}
Eq. (\ref{gf16}) shows that $U\rightarrow\infty$ for
\begin{equation}\label{gf17}
\frac{\|\mathbf{x}-\mathbf{x}_{0}\|}{v}=t-t_{0}
 ~.
\end{equation}
Let us assume that $U$ is measured and found to be very large (i.e., infinite) at the two (measured) space-time points $(\mathbf{x}_{1},t_{1}>t_{0})$ and $(\mathbf{x}_{2},t_{2}>t_{0})$. Then
\begin{equation}\label{gf18}
t_{1}-t_{0}=\frac{\|\mathbf{x}_{1}-\mathbf{x}_{0}\|}{v}~~,~~t_{2}-t_{0}=\frac{\|\mathbf{x}_{2}-\mathbf{x}_{0}\|}{v}
 ~,
\end{equation}
whence the difference
\begin{equation}\label{gf19}
t_{2}-t_{1}=\frac{\|\mathbf{x}_{2}-\mathbf{x}_{0}\|-\|\mathbf{x}_{1}-\mathbf{x}_{0}\|}{v}
 ~.
\end{equation}
The paraxial assumption $|y_{j}-y_{0}|<<|x_{j}-x_{0}|~j=1,2$ entails the approximation (\ref{fgfp2}) whose consequence in  (\ref{gf19}) is
\begin{equation}\label{gf20}
t_{2}-t_{1}\approx\frac{|x_{2}-x_{0}|\left[1+\frac{1}{2}\left(\frac{y_{2}-y_{0}}{x_{2}-x_{0}}\right)^{2}\right]-
|x_{1}-x_{0}|\left[1+\frac{1}{2}\left(\frac{y_{2}-y_{0}}{x_{1}-x_{0}}\right)^{2}\right]}{v}
 ~,
\end{equation}
or neglecting $\frac{1}{2}\left(\frac{y_{j}-y_{0}}{x_{j}-x_{0}}\right)^{2}~;~j=1,2$ and assuming that $x_{j}>x_{0}~;~j=1,2$, we find
\begin{equation}\label{gf20}
v\approx\frac{x_{2}-x_{1}}{t_{2}-t_{1}}
 ~,
\end{equation}
which is the familiar Time-of-flight (TOF), or kinematic, formula for the retrieval of $v$. Note that this retrieval of $v$ is obtained via a measurement procedure in which the coordinates of two space-time points are observed at which the Green's function diverges (compare this to what is written further on in sects. \ref{retv1}-\ref{retv2}).
%%%%%%%%%%%%%%%%%%%%%%%%%%%%%%%%%%%%%%%%%%%%%%%%%%%%%%%%%%%%%%%%%%%%%%%%%%%%%%%
\subsection{The space-time expression of the radiated pressure produced by a more realistic pulse-like line source}\label{rpgf}
%
%%%%%%%%%%%%%%%%%%%%%%%%%%%%%%%%%%%%%%%%%%%
\subsubsection{The pseudo-Ricker pulse spectrum}\label{prps}
It is impossible to produce a Dirac delta pulse source in  practice. An example of something that is closer to a physically-realizable pulse is the pseudo-Ricker pulse whose spectrum is
\begin{equation}\label{prps1}
\mathfrak{h}(\omega)=A\frac{\omega^{\frac{5}{2}}}{4\alpha^{3}\sqrt{\pi}}\exp\left[i\left(\beta\omega-\frac{\pi}{4}\right)-
\frac{\omega^{2}}{4\alpha^{2}}\right]~.
\end{equation}
wherein $A,~\alpha,~\beta$ are real constants.

It is easy to verify that this spectrum satisfies the constraint (\ref{pp15}).
%%%%%%%%%%%%%%%%%%%%%%%%%%%%%%%%%%%%%%%%%%%
\subsubsection{The exact expression of the field radiated by a line source with pseudo-Ricker pulse spectrum}\label{efpr}
We found previously, that the exact expression of the field radiated by a line source is
\begin{equation}\label{efpr1}
U(\mathbf{x},t)=\int_{-\infty}^{\infty}\mathfrak{h}(\omega)\frac{i}{4}
H_{0}^{(1)}(k\|\mathbf{x}-\mathbf{x}_{0}\|)\exp(-i\omega t)d\omega
~,
\end{equation}
which, after the introduction of (\ref{prps1}), becomes
\begin{equation}\label{efpr2}
U(\mathbf{x},t)=\frac{A}{16\alpha^{3}\sqrt{\pi}}
\int_{-\infty}^{\infty}\omega^{\frac{5}{2}}H_{0}^{(1)}\left(\frac{\omega}{v}\|\mathbf{x}-\mathbf{x}_{0}\|\right)\exp
\left\{i\left[\omega(\beta-t)+\frac{\pi}{4}\right]-\frac{\omega^{2}}{4\alpha^{2}}\right\}
d\omega
~.
\end{equation}
The computation of this field at arbitrary points $\mathbf{x}$ will be treated further on.
%%%%%%%%%%%%%%%%%%%%%%%%%%%%%%%%%%%%%%%%%%%
\subsubsection{The far-field of a line source with pseudo-Ricker pulse spectrum}\label{ffpr}
 The introduction of  the far-field approximation (\ref{fgf2}) of the space-frequency field radiated by a line source into (\ref{efpr2}) gives:
\begin{equation}\label{ffpr1}
U(\mathbf{x},t)\approx\int_{-\infty}^{\infty}\mathfrak{h}(\omega)\frac{i}{4}\sqrt{\frac{2}{\pi k\|\mathbf{x}-\mathbf{x}_{0}\|}}\exp\left[\left(ik\|\mathbf{x}-\mathbf{x}_{0}\|-\frac{i\pi}{4}\right)\right]
\exp(-i\omega t)d\omega
~.
\end{equation}
Before going any further, it should be stressed that this operation is fraught with peril, since the far-field approximation can also be interpreted as a high-frequency approximation for reasonable observation point-to-source point distances, and we are introducing this high-frequency approximation into an expression that involves {\it all frequencies, including very small frequencies}. The error one incurs by doing this can be appreciated by inspection of the graphs in sect. \ref{ncspf}.

Be this as it may, we continue the operation to see where it will lead us. Thus, the introduction of (\ref{prps1}) into (\ref{ffpr1}) gives
\begin{equation}\label{ffpr2}
U(\mathbf{x},t)=\frac{A}{16\alpha^{3}\pi}\sqrt{\frac{2v}{|x-x_{0}|}}
\int_{-\infty}^{\infty}\omega^{2}\exp
\left\{i\omega\left[\frac{\|\mathbf{x}-\mathbf{x}_{0}\|}{v}+\beta-t\right]-\frac{\omega^{2}}{4\alpha^{2}}\right\}
d\omega
~,
\end{equation}
%
%%%%%%%%%%%%%%%%%%%%%%%%%%%%%%%%%%%%%%%%%%%
\subsubsection{The paraxial far-field of a line source with pseudo-Ricker pulse spectrum}\label{pffpr}
Assuming (\ref{fgfp1}), and making use of (\ref{fgfp2}), leads to
\begin{equation}\label{pffpr3}
U(\mathbf{x},t)=\frac{A}{16\alpha^{3}\pi}\sqrt{\frac{2v}{|x-x_{0}|}}\mathfrak{J}(\mathbf{x},t)
~,
\end{equation}
wherein
\begin{equation}\label{pffpr4}
\mathfrak{J}(\mathbf{x},t):=\int_{-\infty}^{\infty}\omega^{2}\exp
\left\{-\frac{1}{4\alpha^{2}}\left[\omega^{2}+4i\alpha^{2}\omega(t-\mathfrak{b})\right]\right\}d\omega
~,
\end{equation}
\begin{equation}\label{pffpr5}
\mathfrak{b}(\mathbf{x},t):=\beta+\frac{|x-x_{0}|}{v}+
\frac{1}{2v}\frac{(y-y_{0})^{2}}{|x-x_{0}|}
~.
\end{equation}
But $\omega^{2}+4i\alpha^{2}\omega(t-\mathfrak{b})=[\omega+2i\alpha^{2}(t-\mathfrak{b})]^{2}+4\alpha^{4}(t-\mathfrak{b})^{2}$ so that
\begin{equation}\label{pffpr6}
\mathfrak{J}(\mathbf{x},t)=e^{-\alpha^{2}(t-\mathfrak{b})^{2}}\int_{-\infty}^{\infty}\omega^{2}
\exp\left\{\frac{-[\omega+2i\alpha^{2}(t-\mathfrak{b})]^{2}}{4\alpha^{2}}\right\}d\omega
~.
\end{equation}
With the change of variables $\varpi=\omega+2i\alpha^{2}(t-\mathfrak{b})$, we obtain
\begin{equation}\label{pffpr7}
\mathfrak{J}(\mathbf{x},t)=
e^{-\alpha^{2}(t-\mathfrak{b})^{2}}\Big[\mathfrak{I}_{2}-4i\alpha^{2}(t-\mathfrak{b})\mathfrak{I}_{1}-
4\alpha^{4}(t-\mathfrak{b})^{2}\mathfrak{I}_{0}\Big]
~,
\end{equation}
wherein (Abramowitz \& Stegun, 1968)
\begin{multline}\label{pffpr8}
\mathfrak{I}_{0}=\int_{-\infty}^{\infty}\exp\left(-\frac{\varpi^{2}}{4\alpha^{2}}\right)d\varpi=2\alpha\sqrt{\pi}~~,~~
\mathfrak{I}_{1}=\int_{-\infty}^{\infty}\varpi\exp\left(-\frac{\varpi^{2}}{4\alpha^{2}}\right)d\varpi=0~~,\\
\mathfrak{I}_{2}=\int_{-\infty}^{\infty}\varpi^{2}\exp\left(-\frac{\varpi^{2}}{4\alpha^{2}}\right)d\varpi=4\alpha^{3}\sqrt{\pi}~.
\end{multline}
It follows that
\begin{equation}\label{pffpr10}
U(\mathbf{x},t)=\frac{A}{4}\sqrt{\frac{2v}{\pi |x-x_{0}|}}~
[1-2\alpha^{2}(t-\mathfrak{b})^{2}]e^{-\alpha^{2}(t-\mathfrak{b})^{2}}
~.
\end{equation}
For fixed $\mathbf{x}$, the extrema of $U$ are obtained for $U_{,t}=0$, i.e.,
\begin{equation}\label{pffpr11}
-2\alpha^{2}(t-\mathfrak{b})
\{2+[1-2\alpha^{2}(t-\mathfrak{b})^{2}]\}
e^{-\alpha^{2}(t-\mathfrak{b})^{2}}=0
~,
\end{equation}
the three solutions of which are:
\begin{equation}\label{pffpr12}
t=\tau_{1}=\mathfrak{b}
-\sqrt{\frac{3}{2\alpha^{2}}}~,~~t=\tau_{3}=\mathfrak{b}~,~~
t=\tau_{5}=\mathfrak{b}+\sqrt{\frac{3}{2\alpha^{2}}}
~,
\end{equation}
It turns out that $\tau_{3}$ is the position of the maximum ($=A$) and $\tau_{1},~\tau_{5}$ the positions of the minima, i.e., {\it$\beta$ controls the position of the pulse on the time axis}.

We also notice that $U=0$ when
\begin{equation}\label{pffpr13}
t=\tau_{2}=\mathfrak{b}-\sqrt{\frac{1}{2\alpha^{2}}},~~,
~~t=\tau_{4}=\mathfrak{b}+\sqrt{\frac{1}{2\alpha^{2}}}~,
\end{equation}
 so that $\sqrt{\frac{2}{\alpha^{2}}}$ constitutes a sort of  'width' of the main lobe of the pulse. It ensues that this width decreases as $\alpha$ increases, i.e., {\it $\alpha^{-1}$ controls the pulse width on the time axis}.
%%%%%%%%%%%%%%%%%%%%%%%%%%%%%%%%%%%%%%%%%%%%%%%%%%%%%%%%%%%%%%%%%%%%
\subsection{Methods for retrieving $v$ from the signal}\label{retv}
%
%%%%%%%%%%%%%%%%%%%%%%%%%%%%%%%%%%%%%%%%%%%%%
\subsubsection{Data acquisition}
Suppose that $(\mathbf{x}_{i},~t=t_{ij})$ is any space-time point at which a signal described by (\ref{pffpr10}) exhibits some remarkable feature such as a minimum (at $t=t_{i1},t_{i5}$), a null (at $t=t_{i4}$), or a maximum (at $t_{i3}$). We assume that it is possible to register both the coordinates of $\mathbf{x}_{i}$ and the time $t_{ij}$.

The  general scheme for acquiring the data is as follows:\\
1- emit the pulse at the source point $\mathbf{x}_{0}$,\\
2- at a receiver placed at a point $\mathbf{x}_{1}$, acquire the signal  and note the instant $t_{1j}$ at which the remarkable feature appears,\\
3- at a receiver (with the same characteristics as the previous receiver) placed at another point $\mathbf{x}_{2}$, acquire the signal  and note the instant $t_{2j}$ at which the same remarkable feature as previously  appears.\\
Thus, the data consists of the two pairs: $(\mathbf{x}_{1},t_{1j})$ and $(\mathbf{x}_{2},t_{2j})$.

Note that the characteristics of the source, i.e., its location $\mathbf{x}_{0}$, $\alpha$ and $\beta$ are assumed  not to change between the two measurement space-time points.
%%%%%%%%%%%%%%%%%%%%%%%%%%%%%%%%%%%%%%%%%%%%%%%%%%%%%%%%%%%%%%%%%%%%
\subsubsection{Method for retrieving $v$ employing $\tau_{1}$}\label{retv1}
Due to the assumed constancy of $\mathbf{x}_{0}$, $\alpha$ and $\beta$,  (\ref{pffpr11}-(\ref{pffpr12}) tell us that:
\begin{equation}\label{pffpr14}
t_{1}=t_{11}=\mathfrak{b}_{1}-\sqrt{\frac{3}{2\alpha^{2}}}~~,~~t_{2}=t_{21}=\mathfrak{b}_{2}-\sqrt{\frac{3}{2\alpha^{2}}}
~,
\end{equation}
wherein
\begin{equation}\label{pffpr15}
\mathfrak{b}_{i}=\beta+\frac{|x_{i}-x_{0}|}{v}+
\frac{1}{2v}\frac{(y_{i}-y_{0})^{2}}{|x_{i}-x_{0}|}~~,~~i=1,2
~,
\end{equation}
and $\mathbf{x}_{i}=(x_{i},y_{i})$. By subtraction we obtain:
\begin{equation}\label{pffpr16}
t_{2}-t_{1}=\mathfrak{b}_{2}-\mathfrak{b}_{1}=\frac{|x_{2}-x_{0}|}{v}+
\frac{1}{2v}\frac{(y_{2}-y_{0})^{2}}{|x_{2}-x_{0}|}-\frac{|x_{1}-x_{0}|}{v}-
\frac{1}{2v}\frac{(y_{1}-y_{0})^{2}}{|x_{1}-x_{0}|}
~,
\end{equation}
whence
\begin{equation}\label{pffpr17}
v=\frac{|x_{2}-x_{0}|+\frac{(y_{2}-y_{0})^{2}}{|x_{2}-x_{0}|}-|x_{1}-x_{0}|-
\frac{(y_{1}-y_{0})^{2}}{|x_{1}-x_{0}|}}{t_{2}-t_{1}}
~,
\end{equation}
This expression shows that  $v$ is retrieved under the condition that we know beforehand the location $\mathbf{x}_{0}$. There exist two manners for partially or fully avoiding this constraint. The first is to assume that we are able to place the receivers on the same $y$ axis as the source, i.e., $y_{j}=y_{0}~;~j=1,2$, in which case
\begin{equation}\label{pffpr18}
v=\frac{|x_{2}-x_{0}|-|x_{1}-x_{0}|}{t_{2}-t_{1}}
~,
\end{equation}
or, assuming that $x_{j}>x_{0}~;~j=1,2$,
\begin{equation}\label{pffpr19}
v=\frac{x_{2}-x_{1}}{t_{2}-t_{1}}
~,
\end{equation}
which is the familiar TOF formula for the retrieval of $v$.

The second--easier to realize--manner is to assume that $\frac{(y_{j}-y_{0})^{2}}{|x_{j}-x_{0}|}<<|x_{j}-x_{0}|~;~j=1,2$, in which case we again obtain (\ref{pffpr18}) or (\ref{pffpr19}).

The same procedure can be applied to $t_{5}$ data  and gives rise to the same result.

Note that the outlined retrieval procedure is {\it kinematic} in nature, since only the times of arrival of certain features of the radiated signal are chosen as data, not the amplitude of the signals (as in a {\it dynamic} procedure).
%%%%%%%%%%%%%%%%%%%%%%%%%%%%%%%%%%%%%%%%%%%%%%%%%%%%%%%%%%%%%%%%%%%%
\subsubsection{Method for retrieving $v$ employing $\tau_{2}$}
Due to the assumed constancy of $\mathbf{x}_{0}$, $\alpha$ and $\beta$,  (\ref{pffpr13}) tells us that:
\begin{equation}\label{pffpr20}
t_{1}=t_{12}=\mathfrak{b}_{1}-\sqrt{\frac{1}{2\alpha^{2}}}~~,~~t_{2}=t_{22}=\mathfrak{b}_{2}-\sqrt{\frac{1}{2\alpha^{2}}}
\end{equation}
By subtraction we again obtain (\ref{pffpr16}), wherein $t_{i2}$ replaces $t_{i1}$), so that all the previous formulae (\ref{pffpr17})-(\ref{pffpr19}) are again applicable, including the TOF formula for the retrieval of $v$.

The same procedure can be applied to $t_{4}$ data   and gives rise to the same result.
%%%%%%%%%%%%%%%%%%%%%%%%%%%%%%%%%%%%%%%%%%%%%%%%%%%%%%%%%%%%%%%%%%%%
\subsubsection{Method for retrieving $v$ employing $\tau_{3}$}\label{retv2}%\label{mrvt0}%
Using the maximum of the pulse as its remarkable feature is commonplace in parameter (such as $v$) retrieval practice. Due to the assumed constancy of $\mathbf{x}_{0}$, $\alpha$ and $\beta$,  (\ref{pffpr12}) tells us that:
\begin{equation}\label{pffpr21}
t_{1}=t_{13}=\mathfrak{b}_{1}~~,~~t_{2}=t_{23}=\mathfrak{b}_{2}
\end{equation}
By subtraction we again obtain (\ref{pffpr16}), , wherein $t_{i3}$ replaces $t_{i3}$), so that all the previous formulae (\ref{pffpr17})-(\ref{pffpr19}) are again applicable, including the TOF formula for the retrieval of $v$.
%%%%%%%%%%%%%%%%%%%%%%%%%%%%%%%%%%%%%%%%%%%
\subsection{Computation of the space-time field radiated by a line source with pseudo-Ricker pulse spectrum at arbitrary $\mathbf{x}$}\label{efpr}
We found previously, that  the exact expression of the field radiated by a line source is
\begin{equation}\label{efpr1}
U(\mathbf{x}),t)=2\Re\int_{0}^{\infty}\mathfrak{h}(\omega)\frac{i}{4}
H_{0}^{(1)}(k\|\mathbf{x}-\mathbf{x}_{0}\|)\exp(-i\omega t)d\omega
~,
\end{equation}
which, after the introduction of (\ref{prps1}), relative to a pseudo-Ricker pulse,  becomes
\begin{equation}\label{efpr3}
U(\mathbf{x}),t)=
\frac{A}{18\alpha^{3}\sqrt{\pi}}\Re\left\{\mathfrak{I}_{1}(\mathbf{x},t)e^{i\frac{\pi}{4}}\right\}+
\frac{A}{18\alpha^{3}\sqrt{\pi}}\Re\left\{\mathfrak{I}_{2}(\mathbf{x},t)e^{i\frac{\pi}{4}}\right\}=
U_{1}(\mathbf{x}),t)+U_{2}(\mathbf{x}),t)
~,
\end{equation}
wherein:
\begin{equation}\label{efpr5}
\mathfrak{I}_{1}(\mathbf{x},t)=\int_{0}^{\varepsilon}\omega^{\frac{5}{2}}H_{0}^{(1)}(k\|\mathbf{x}-\mathbf{x}_{0}\|)
\exp\left[i(\beta-t)-\frac{\omega^{2}}{4\alpha^{2}}\right]
d\omega~,
\end{equation}
\begin{equation}\label{efpr6}
\mathfrak{I}_{2}(\mathbf{x},t)=\int_{\varepsilon}^{\infty}\omega^{\frac{5}{2}}H_{0}^{(1)}(k\|\mathbf{x}-\mathbf{x}_{0}\|)
\exp\left[i(\beta-t)-\frac{\omega^{2}}{4\alpha^{2}}\right]
d\omega~.
\end{equation}
and $\varepsilon$ is a small quantity. Let us first consider $\mathfrak{I}_{1}$, wherein $Z=\frac{\omega}{v}\|\mathbf{x}-\mathbf{x}_{0}\|\in [0,\varepsilon]$ Under the assumption $0<Z<<1~;~\forall Z\in [0,\varepsilon]$, we can employ the approximation (Abramowitz \& Stegun, 1968)
\begin{equation}\label{efpr7}
H_{0}^{(1)}(Z)\approx 1+\frac{2i}{\pi}\left(\ln \frac{Z}{2}+\gamma\right)
~,
\end{equation}
wherein $\gamma=0.5772156649$. We place two further constraints on $\varepsilon$: $|(\beta-t)\varepsilon<<1$ and $\frac{\varepsilon^{2}}{4\alpha^{2}}<<1$. The first constraint authorizes the approximation $\exp(i\omega(\beta-t))\sim 1$ and the second constraint authorizes $\exp(-\omega^{2}/4\alpha^{2})\sim 1$, so that
\begin{equation}\label{efpr8}
\mathfrak{I}_{1}(\mathbf{x},t)\sim\int_{0}^{\varepsilon}\omega^{\frac{5}{2}}\left[1+\frac{2i}{\pi}\left(\ln \frac{\omega\|\mathbf{x}-\mathbf{x}_{0}\|}{2v}+\gamma\right)\right]
d\omega=\mathfrak{I}_{11}(\mathbf{x},t)+\mathfrak{I}_{12}(\mathbf{x},t)~.
\end{equation}
The two sub-integrals can be evaluated explicitly:
\begin{equation}\label{efpr9}
\mathfrak{I}_{11}(\mathbf{x},t)=\left(1+\frac{2i\gamma}{\pi}\right)\int_{0}^{\varepsilon}\omega^{\frac{5}{2}}d\omega=
\left(1+\frac{2i\gamma}{\pi}\right)\frac{2}{7}~\varepsilon^{\frac{7}{2}}~,
\end{equation}
\begin{equation}\label{efpr10}
\mathfrak{I}_{12}(\mathbf{x},t)=\left(\frac{2i}{\pi}\right)\int_{0}^{\varepsilon}\omega^{\frac{5}{2}}\ln \left( \frac{\omega\|\mathbf{x}-\mathbf{x}_{0}\|}{2v}\right)d\omega=
\left(\frac{4i}{7\pi}\right)~\varepsilon^{\frac{7}{2}}\left[\ln \left( \frac{\varepsilon\|\mathbf{x}-\mathbf{x}_{0}\|}{2v}\right)-\frac{2}{7} \right]~,
\end{equation}
Although  $I_{11}$, $I_{12}$, and thus $I_{1}$, $\rightarrow 0$ as $\varepsilon\rightarrow 0$, it is easy, as we have done, to compute them via (\ref{efpr9})-(\ref{efpr10}), with  $\varepsilon$ chosen so as to satisfy the three constraints: $0<Z<<1~;~\forall Z\in [0,\varepsilon]$, $|(\beta-t)\varepsilon<<1$ and $\frac{\varepsilon^{2}}{4\alpha^{2}}<<1$. The remaining integral, i.e., $I_{2}$ can be performed by any (we chose the Simpson) numerical quadrature scheme.

In this manner, we are able to compute the  values of the space-time field radiated by the pseudo-Ricker pulse line source to any desired degree of precision. This is done in sect. \ref{nfff}  and  comparisons are made therein with the corresponding Green's function signals and far-field paraxial signals.
%\newpage
%%%%%%%%%%%%%%%%%%%%%%%%%%%%%%%%%%%%%%%%%%%%%%%%%%%%%%%%%%%%%
\subsection{Paraxial near and far field signals}\label{nfff}
In  the following figures, we have taken: $x_{0}=y_{0}=y=0$, $A=1$, $\beta=10s$, $v=1500~m/s$.
\begin{figure}
[ht]
\begin{center}
\includegraphics[scale=0.5]{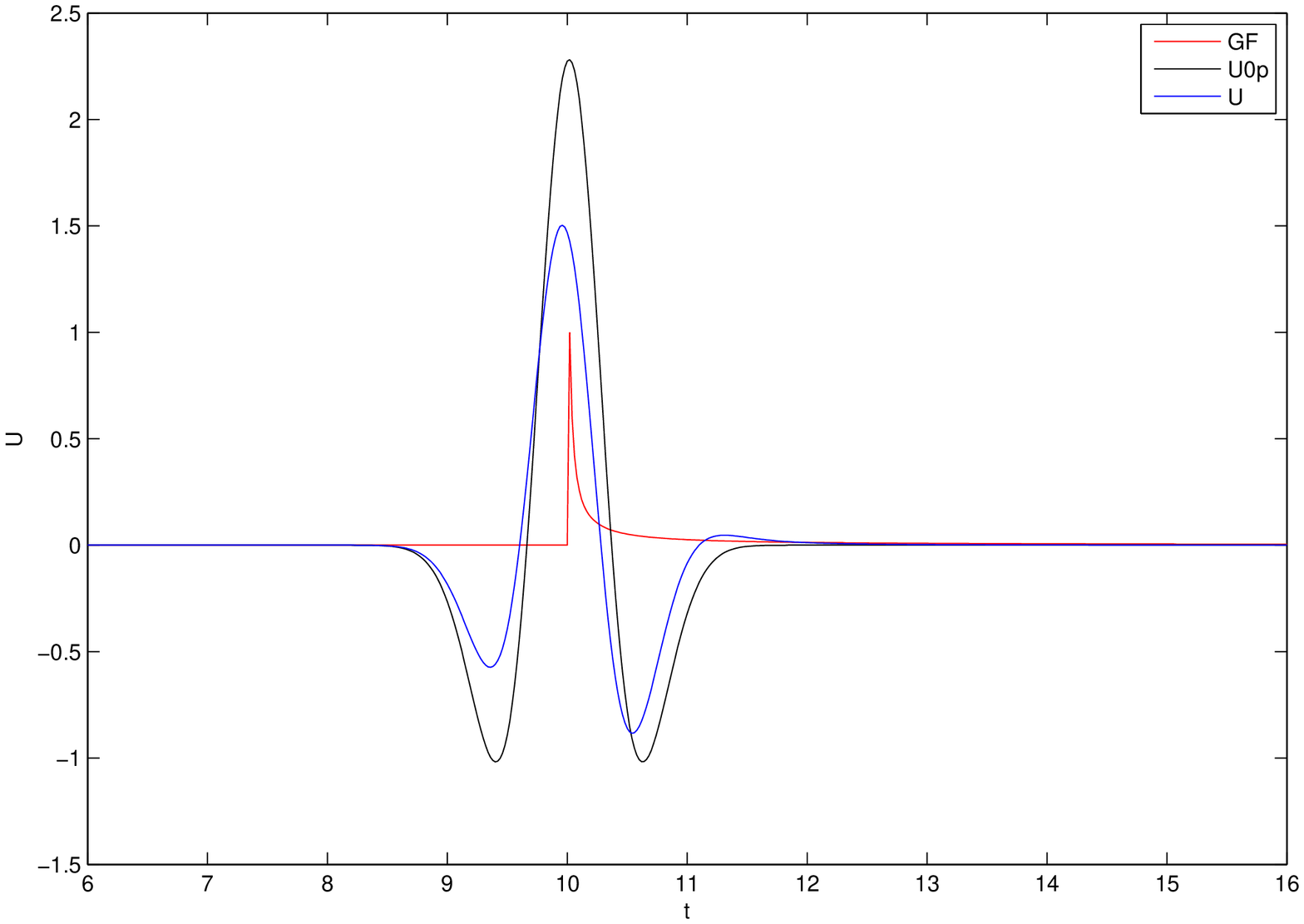}
\caption{Pressure signal $U$(a.u.) as a function of $t(s)$ at a fixed near-field location $x=25 m$. The red curve is the Green's function. The black curve is the paraxial far-field expression of the signal. The blue curve is the exact expression of the signal. $\alpha=2~Hz$.} \label{gf-3}
\end{center}
\end{figure}
\begin{figure}
[ht]
\begin{center}
\includegraphics[scale=0.5]{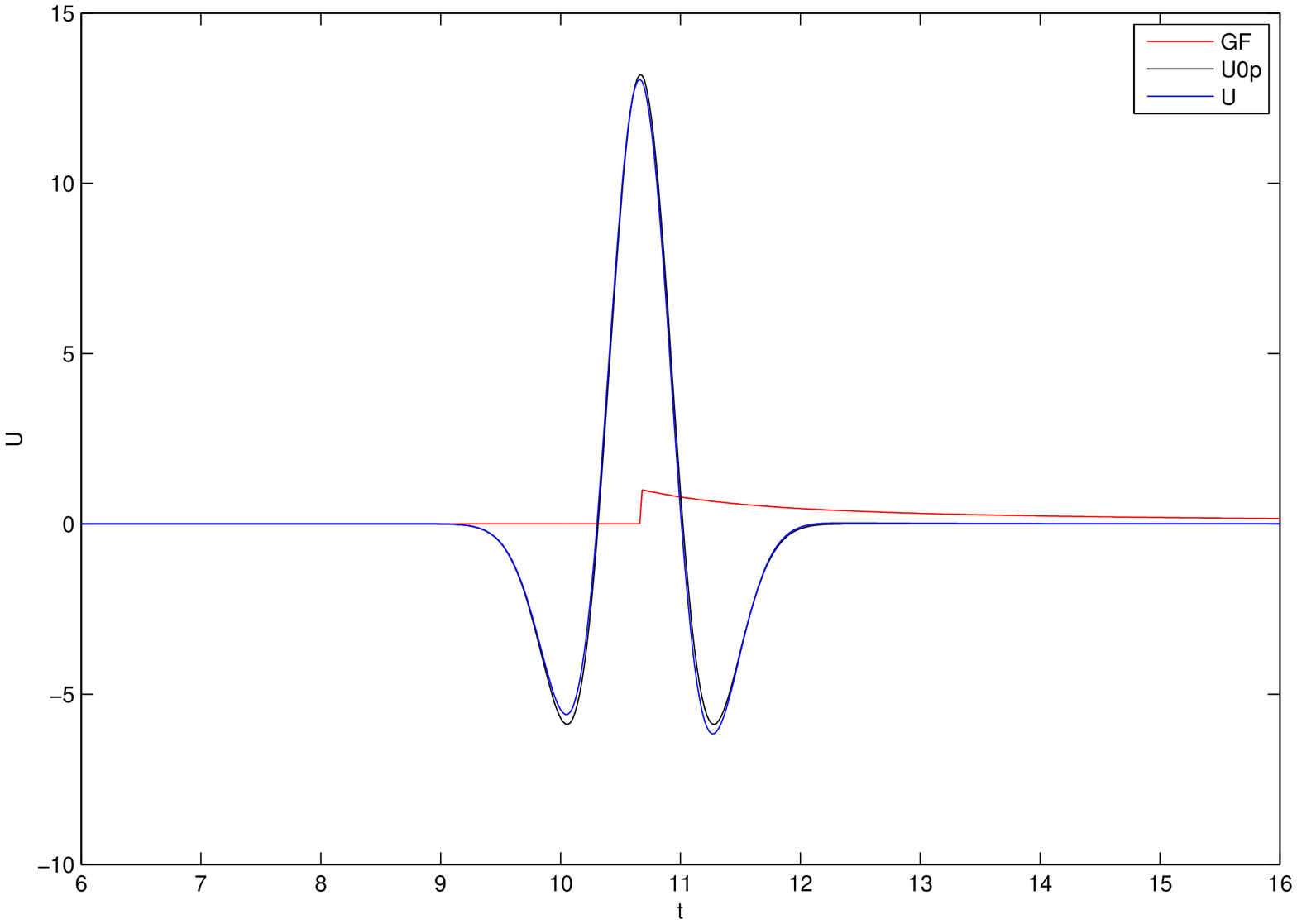}
\caption{Pressure signal $U$(a.u.) as a function of $t(s)$ at a fixed far-field location $x=1000 m$. The red curve is the Green's function. The black curve is the paraxial far-field expression of the signal. The blue curve is the exact expression of the signal. $\alpha=2~Hz$.} \label{gf-3}
\end{center}
\end{figure}
\clearpage
\newpage
%%%%%%%%%%%%%%%%%%%%%%%%%%%%%%%%%%%%%%%%%%%%%%%%%%%%%%%%%%%%%%%%%%%%%%%%%%%%%%%%%%%%%%%%%%%%%%%%%%%%%%%%%%%%%%%%%%%%%%%%
\subsection{Numerical applications of the TOF method using  the exact  expression of the field generated by a pseudo-Ricker pulse}\label{mrvt0r}
We employed the  TOF retrieval scheme outlined in sect. \ref{retv}, based on the formula $v=\frac{x_{2}-x_{1}}{t_{2}-t_{1}}$, with $t_{1},~t_{2}$ the instants of occurrence of the maximum of the received pulse when the latter is computed by the exact expression of the signal as outlined in sect. \ref{efpr}. The true fluid velocity was $v=1500~m/s$, whereas the source was at $x_{0}=y_{0}=0$ and emitted a pseudo-Ricker pulse whose maximum was at $\beta=10~s$. The receivers were placed at different locations along the $x-$ axis.
%%%%%%%%%%%%%%%%%%%%%%%%%%%%%%%%%%%%%%%%%%%%%%%%%%%%%%%%%%%%%%%%%%%%%%%%%%%%%%%%%%%%%%%%%%%%%%%%%%%%%%%%%%%%%%%%%%%%%%%%
\subsubsection{First numerical application}\label{mrvt0r1}

\begin{figure}
[ht]
\begin{center}
\includegraphics[scale=0.4]{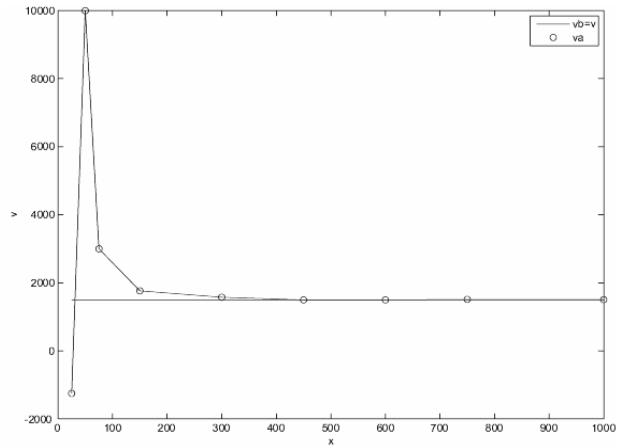}
\caption{Retrieved $v$ versus $x=x_{2}$ when $x_{1}=x_{0}=0$. The line labeled $vb$ is the true velocity. The curve and circles labeled $va$ is the retrieved velocity obtained from the instant of occurrence of the maximum of the received pulse when the latter is computed by the exact expression of the signal. Case $\alpha=3~Hz$.} \label{mrvt0r-1}
\end{center}
\end{figure}

\begin{figure}
[ht]
\begin{center}
\includegraphics[scale=0.4]{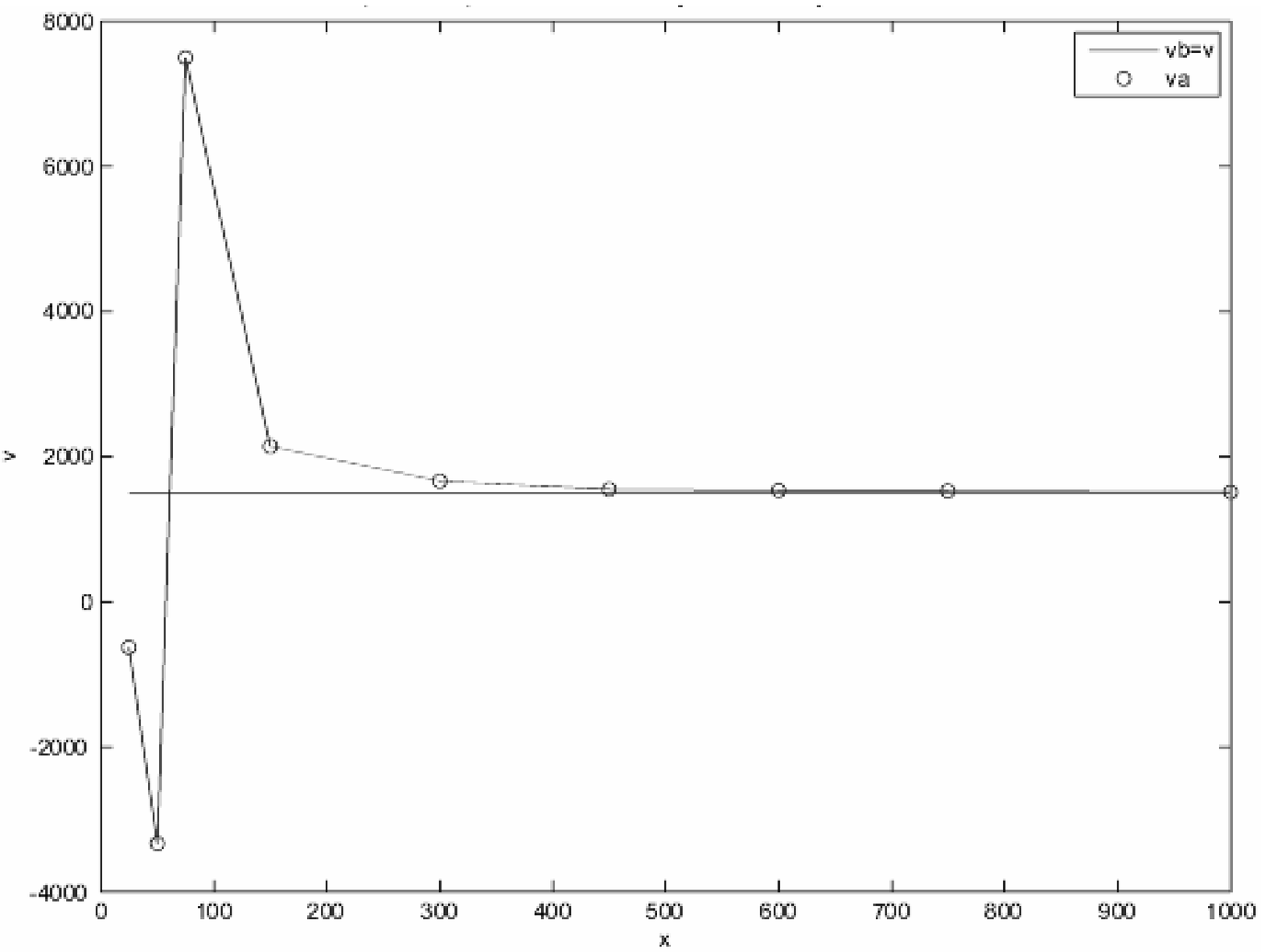}
\caption{Retrieved $v$ versus $x=x_{2}$ when $x_{1}=x_{0}=0$. The line labeled $vb$ is the true velocity. The curve and circles labeled $va$ is the retrieved velocity obtained from the instant of occurrence of the maximum of the received pulse when the latter is computed by the exact expression of the signal.  Case $\alpha=2~Hz$.} \label{mrvt0r-2}
\end{center}
\end{figure}

\begin{figure}
[ht]
\begin{center}
\includegraphics[scale=0.4]{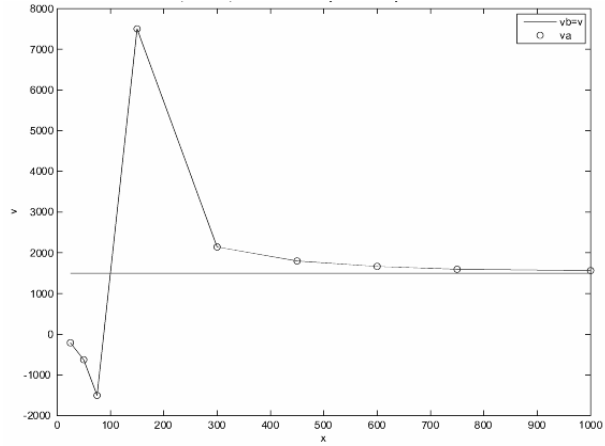}
\caption{Retrieved $v$ versus $x=x_{2}$ when $x_{1}=x_{0}=0$. The line labeled $vb$ is the true velocity. The curve and circles labeled $va$ is the retrieved velocity obtained from the instant of occurrence of the maximum of the received pulse when the latter is computed by the exact expression of the signal.  Case $\alpha=1~Hz$.} \label{mrvt0r-3}
\end{center}
\end{figure}

In the first application, we fixed the first observation point at the source, i.e., $x_{1}=x_{0}$ and $\tau_{1}=\beta$, and varied the location of the second observation point along the $x-$ axis, so that $v=\frac{x_{2}-0}{\tau_{2}-\beta}$. The results, constituting plots of the retrieved $v$ versus the $x=x_{2}$ coordinate  are depicted in figs. \ref{mrvt0r-1}-\ref{mrvt0r-3} for three different pseudo-Ricker pulses which differ by the parameter $\alpha$ (recall that the smaller is $\alpha$, the wider is the pulse).
\clearpage
\newpage
The upshot of these results is that if the receiver is close to the source, the TOF method applied to real data (embodied in the exact expression of the signal) can lead to enormous errors in the retrieved values of $v$, this being all the more true, the smaller is $\alpha$.

Thus, for a chosen fixed $\alpha$, to obtain a retrieved $v$ with acceptable error, one must place the receiver all the farther from the source than $\alpha$ is smaller. On the other hand, for a fixed receiver-to-source distance $|x-x_{0}|$, to obtain a retrieved $v$ with acceptable error, one must choose the pulse width to be all the larger the smaller is $|x-x_{0}|$.

A remarkable result of these computations relative to  TOF retrieval of $v$ applied to exact data, is that  one  generally obtains (except for very small $|x-x_{0}|$ and/or small $\alpha$)   {\it a retrieved velocity that is larger than the true velocity}.  In fact, the explanation of these large differences in the near-field zone lies in the fact that we employ the wrong model (constituted by the TOF scheme, which, strictly speaking, applies only in the far-field paraxial region zone) to invert  the right data (here, that which results from the computation of the exact signal).
%\clearpage
%\newpage
%%%%%%%%%%%%%%%%%%%%%%%%%%%%%%%%%%%%%%%%%%%%%%%%%%%%%%%%%%%%%%%%%%%%%%%%%%%%%%%%%%%%%%%%%%%%%%%%%%%%%%%%%%%%%%%%%%%%%%%%
\subsubsection{Second numerical application}\label{mrvt0r2}
\begin{figure}
[ht]
\begin{center}
\includegraphics[scale=0.4]{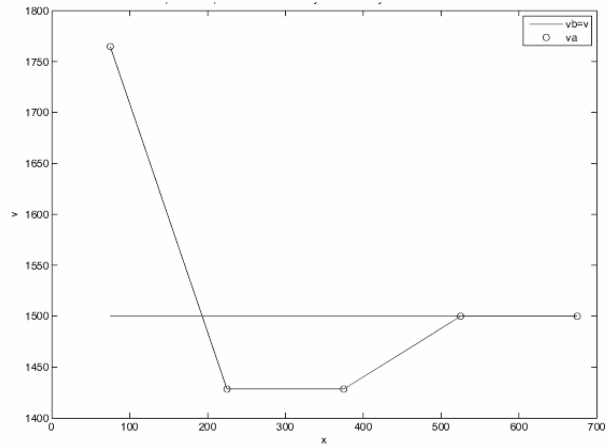}
\caption{Retrieved $v$ versus $x=\frac{x_{2}+x_{1}}{2}$ when $x_{2}-x_{1}=150~m$. The line labeled $vb$ is the true velocity. The curve and circles labeled $va$ is the retrieved velocity obtained from the instant of occurrence of the maximum of the received pulse when the latter is computed by the exact expression of the signal. Case $\alpha=3~Hz$.} \label{mrvt0r-4}
\end{center}
\end{figure}

\begin{figure}
[ht]
\begin{center}
\includegraphics[scale=0.4]{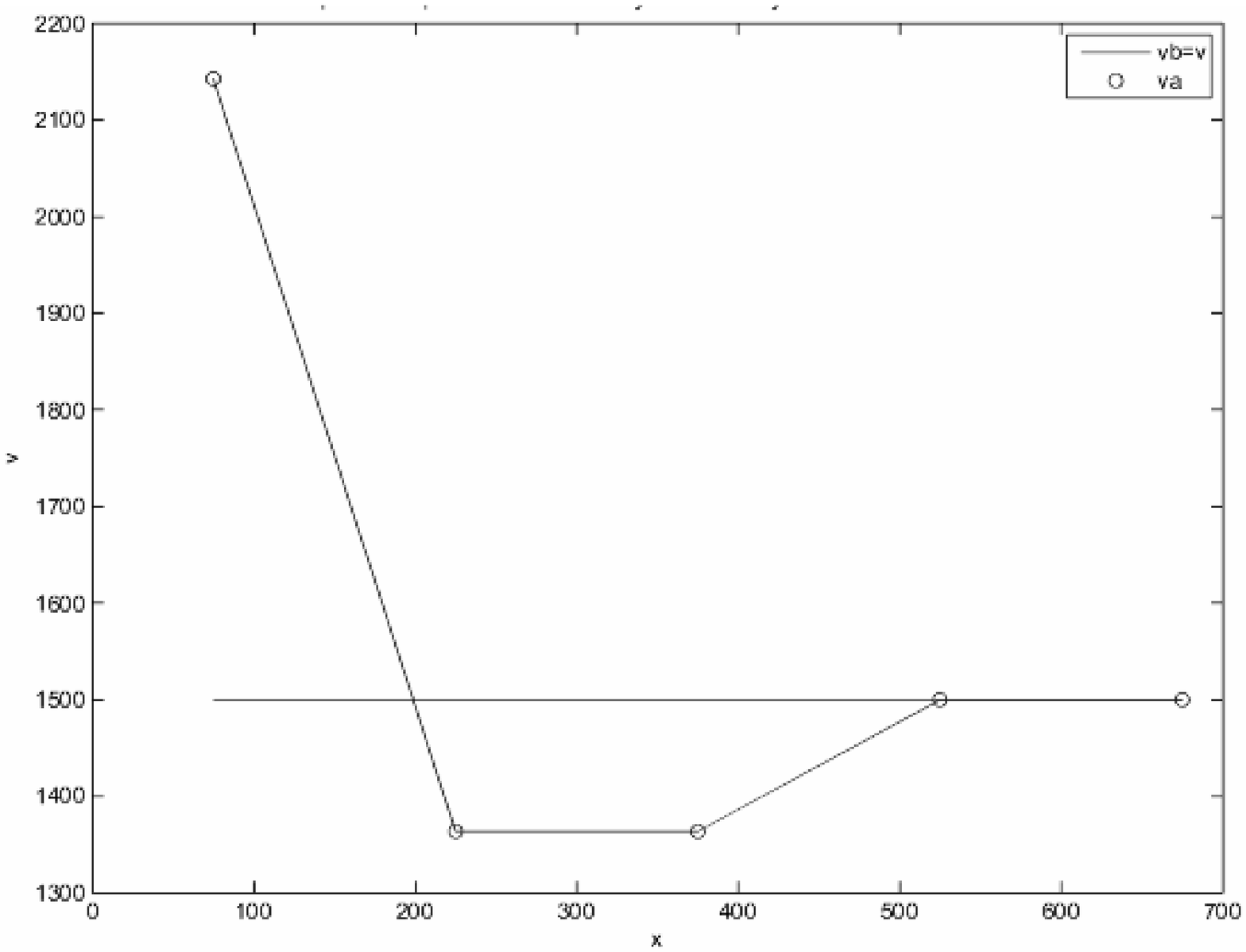}
\caption{Retrieved $v$ versus $x=\frac{x_{2}+x_{1}}{2}$ when $x_{2}-x_{1}=150~m$. The line labeled $vb$ is the true velocity. The curve and circles labeled $va$ is the retrieved velocity obtained from the instant of occurrence of the maximum of the received pulse when the latter is computed by the exact expression of the signal. Case $\alpha=2~Hz$.} \label{mrvt0r-5}
\end{center}
\end{figure}

\begin{figure}
[ht]
\begin{center}
\includegraphics[scale=0.4]{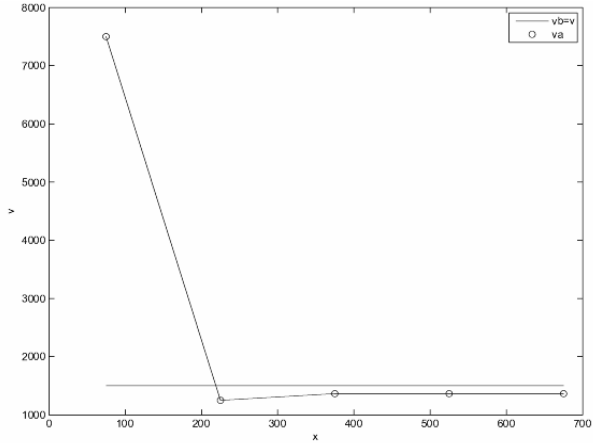}
\caption{Retrieved $v$ versus $x=\frac{x_{2}+x_{1}}{2}$ when $x_{2}-x_{1}=150~m$. The line labeled $vb$ is the true velocity. The curve and circles labeled $va$ is the retrieved velocity obtained from the instant of occurrence of the maximum of the received pulse when the latter is computed by the exact expression of the signal. Case $\alpha=1~Hz$.} \label{mrvt0r-6}
\end{center}
\end{figure}

In the second application, we varied both $x_{1}$ and $x_{2}$ (while keeping $y=0$) so as to have their difference fixed at the value $d=150~m$.  The results, constituting plots of the retrieved $v$ versus the $x=\frac{x_{2}+x_{1}}{2}$ coordinate  are depicted in figs. \ref{mrvt0r-4}-\ref{mrvt0r-6} for three different pseudo-Ricker signals (differentiated by the parameter $\alpha$).
\clearpage
\newpage
We observe in these figures  that, now, for the most part, {\it the retrieved $v$ are smaller than their true value $v=1500~m/s$}. Otherwise, the same comments as previously apply here as well.
%%%%%%%%%%%%%%%%%%%%%%%%%%%%%%%%%%%%%%%%%%%%%%%%%%%%%%%%%%%%%%%%%%%%%%%%%%%%%%%%%%%%%%%%%%%%%%%%%%%%%%%%%%%%%%%%%%%%%%%%
\subsection{Taking account of uncertainty in $\beta$ during the retrieval of $v$ via the paraxial far-field approximation of the signal}\label{tspffa}
Let us consider the case in which the first measurement of the position of the pulse maximum is made at the source and gives rise to the data $(\mathbf{x}_{1}:=\mathbf{x}_{0},~t_{1}:=t_{0}=\beta)$, and the second measurement of the position of the pulse maximum gives rise to the data $(\mathbf{x}_{2}:=\mathbf{x},~t_{2}:=t)$. We now focus our attention on the retrieval of $v$ via the data collected at the second receiver, assuming that: i) there is some uncertainty in $\beta$, and ii) it is legitimate to employ the paraxial far-field approximation of the signal.
%%%%%%%%%%%%%%%%%%%%%%%%%%%%%%%%%%%%%%%%%%%%%%%%%%%%%%%%%%%%%%%%%%%%%%%%%%%%%%%%%%%%%%%%%%%%%%%%%%%%%%%%%%%%%%%%%%%%%%%%
\subsubsection{Retrieval of $v$ by the TOF formula taking into account uncertainty of the pulse emission instant}\label{utof}
At present, we make the additional assumption: that it is legitimate to employ the formula for the position of the maximum of the signal to which the paraxial far-field approximation of the signal leads.

According to this formula (see (\ref{pffpr12})),  $t$ is related to $x$ and $v$ in the true signal by
\begin{equation}\label{utof1}
    t=\mathfrak{b}=\beta+\frac{|x-x_{0}|}{v}+\frac{1}{2}\frac{(y-y_{0})^{2}}{|x-x_{0}|}~.
\end{equation}
To make things simple, we assume that $y=y_{1}=y_{0}$, and $x>x_{0}$, so that (with $d=x-x_{0}$),
\begin{equation}\label{utof2}
    t=\beta+\frac{d}{v}~.
\end{equation}
So much for the data. In order to recover $v$, we must employ a so-called {\it retrieval model}, and to do this, we once again rely on the paraxial far-field approximation of the signal received at $(x,t)$
\begin{equation}\label{utof3}
    t=\beta'+\frac{d}{v'}~.
\end{equation}
wherein $\beta$ has been replaced by $\beta'$ to express the assumed uncertainty  concerning the instant of occurrence of the maximum of the pulse at the position of the source. This uncertainty necessarily leads to an error in the retrieval of the velocity, expressed by the replacement of $v$ by $v'$ in the retrieval model.

From these two formulae, we easily deduce
\begin{equation}\label{utof4}
    v'=\frac{x}{\beta-\beta'+\frac{d}{v}}~,
\end{equation}
or, equivalently,
\begin{equation}\label{utof5}
   \frac{v'}{v}=1+\frac{\beta'-\beta}{t-\beta}~.
\end{equation}
The latter equation shows that if, as is usual,  $t>\beta'$, the retrieved velocity $v'$ is greater than the true velocity $v$ when $\beta'>\beta$. Moreover, the retrieved velocity $v'$ is smaller than the true velocity $v$ when $\beta'<\beta$. If, as is not usually the case, we have perfect knowledge of $\beta$, we find $v'=v$ which means that we recover the true value of the velocity.

An interesting feature of this result is that it is independent of the pulse shape (the latter being controlled by the parameter $\alpha$), but this feature is probably specific to the paraxial far-field approximation model of the signal.

Another interesting feature of this result is that it shows that the error of $v'$ with respect to $v$ decreases as $d$ (and therefore $t$) is increased, which explains why it is important to place the receiver as far as possible from the source.

These features will become apparent in the graphs of the sect. \ref{ncts}.
%%%%%%%%%%%%%%%%%%%%%%%%%%%%%%%%%%%%%%%%%%%%%%%%%%%%%%%%%%%%%%%%%%%%%%%%%%%%%%%%%%%%%%%%%%%%%%%%%%%%%%%%%%%%%%%%%%%%%%%%
\subsubsection{Full-wave retrieval of $v$ by the paraxial far-field approximation of the signal taking into account uncertainty of the pulse emission instant}\label{upffa}
 At present, we no longer rely on the formula (\ref{pffpr12}) for the position of the maximum of the signal, but rather make use of the expression (\ref{pffpr10}) of the signal itself, since the data is now  (a part of) the signal rather than the position of one of its remarkable features. We assume that this signal is acquired over the temporal interval $[t_{b},t_{e}]$.

Rather than actually carry out the measurement experiment, we {\it simulate} this experiment using the paraxial far-field model as the device for obtaining the {\it synthetic data}. This data signal is of the form:
\begin{equation}\label{upffa1}
U(\mathbf{x},t)=\frac{A}{4}\sqrt{\frac{2v}{\pi |x-x_{0}|}}~
[1-2\alpha^{2}(t-\mathfrak{b})^{2}]e^{-\alpha^{2}(t-\mathfrak{b})^{2}}~~;~~t\in[t_{b},t_{e}]
~,
\end{equation}
wherein
\begin{equation}\label{upffa2}
    \mathfrak{b}=\beta+\frac{|x-x_{0}|}{v}+\frac{1}{2}\frac{(y-y_{0})^{2}}{|x-x_{0}|}~.
\end{equation}
and $\mathbf{x}_{0},t,v,A,\alpha,\beta$ are the {\it true} values of space-time position and source characteristics.

To make things simple, we assume that the retrieval model also appeals to the paraxial far-field approximation of the signal, and, that amongst the parameters, all but the equivalent of $\beta$ are exactly known. The retrieval model signal is then
\begin{equation}\label{upffa3}
U'(\mathbf{x},t)=\frac{A}{4}\sqrt{\frac{2v'}{\pi |x-x_{0}|}}~
[1-2\alpha^{2}(t-\mathfrak{b}')^{2}]e^{-\alpha^{2}(t-\mathfrak{b}')^{2}}~~;~~t\in[t_{b},t_{e}]
~,
\end{equation}
wherein
\begin{equation}\label{upffa4}
    \mathfrak{b}'=\beta'+\frac{|x-x_{0}|}{v'}+\frac{1}{2}\frac{(y-y_{0})^{2}}{|x-x_{0}|}~.
\end{equation}
We note, that as in sect. \ref{utof}, these formulae express the fact  that the  uncertainty of $\beta$ necessarily entails a retrieved value $v'$ of the velocity which is different from the true  velocity $v$.

 Our procedure for actually retrieving $v$,  different from the one of sect. \ref{utof}, is increasingly-employed in the non-destructive testing and geophysical communities and termed {\it full-wave inversion} (FWI) (Virieux \ Operto, 2009). The idea is to vary $v'$ within a a finite search interval $\mathfrak{I}\in\mathbb{R}$, generate a  function of the discrepancy between $U$ and $U'$ for each $v'$, and associate the retrieved velocity with that $v'$ for which the discrepancy function attains a global minimum. Our discrepancy function is
\begin{equation}\label{upffa5}
K(v',\beta',v,\beta,t_{b},t_{e})=\frac{\frac{1}{t_{e}-t_{b}}\int_{t_{b}}^{t_{e}}[U(\mathbf{x,t})-U'(\mathbf{x},t)]^{2}dt}
{\frac{1}{t_{e}-t_{b}}\int_{t_{b}}^{t_{e}}[U(\mathbf{x},t)]^{2}dt}~.
\end{equation}
Note that $K$ depends explicitly on $t_{b},t_{e}$. Normally speaking, it also depends on the way the integrals are computed, but, we assume that this can be done numerically with any desired amount of precision.

The (global) minimum of the discrepancy functional is found for $v'=v''$, i.e.,
\begin{equation}\label{ecf4}
    v''(\beta',v,\beta,t_{b},t_{e})=\arg\min_{\mathfrak{I}\subset \mathbb{R}}K(v',\beta',v,\beta,t_{b},t_{e})~,
\end{equation}
wherein $\mathfrak{I}$ is the set of all the values of  $v'$ during a given minimum search process.

Note that  $v''$ depends: a) explicitly on the choices of the prior $\beta'$, the search interval $\mathfrak{I}$, the portion $[t_{b},t_{e}]$ of the signal  employed in the retrieval operation, and b) implicitly on $\alpha,\beta,~v,\mathbf{x}_{0},\mathbf{x}$.

The $arg$ symbol in front of the $min$ symbol means that the actual value of the minimum of $K$ is irrelevant other than the requirement that it be a {\it global minimum} (this meaning that it is possible that $K$ exhibits other so-called local minima, but the one of interest is the deepest, so called global minimum).

It follows that
\begin{equation}\label{ecf9}
K(v'=v,\beta'=\beta,v,\beta,t_{b},t_{e})=0~,
\end{equation}
since the mathematical models underlying $U$ and $U'$ are identical and thus give rise to the same signals for $\beta'=\beta$ and $v'=v$. This implies that when $\beta'=\beta$ (i.e., the prior $\beta$ is exact),  the result of the inversion will lead to at least one solution (Wirgin, 2004) $v''=v$ which happens to be the true solution $v$.
%%%%%%%%%%%%%%%%%%%%%%%%%%%%%%%%%%%%%%%%%%%%%%%%%%%%%%%%%%%%%%%%%%%%%%%%%%%%%%%%%%%%%%%%%%%%%%%%%%%%%%%%%%%%%%%%%%%%%%%%
\subsubsection{Numerical comparison of the two schemes for the retrieval of sound in pure water}\label{ncts}
In the following figures, we compare the results of the retrieval of $v$ by the two previously-described schemes when there is some uncertainty in the instant $t_{0}=\beta$ of the maximum of the initial pseudo-Ricker pulse.

In all these figures, we have taken: $x_{0}=y_{0}=y=0$, $A=1$, $\beta=10~s$, and we vary the position $x$ of the receiver along the $x$-axis.
The true value of the disturbance is assumed to be $v=1500~m/s$, which is close to the accepted value of the velocity of acoustic bulk waves in pure water at room temperature.
\clearpage
\newpage
\begin{figure}
[ptb]
\begin{center}
\includegraphics[scale=0.4]{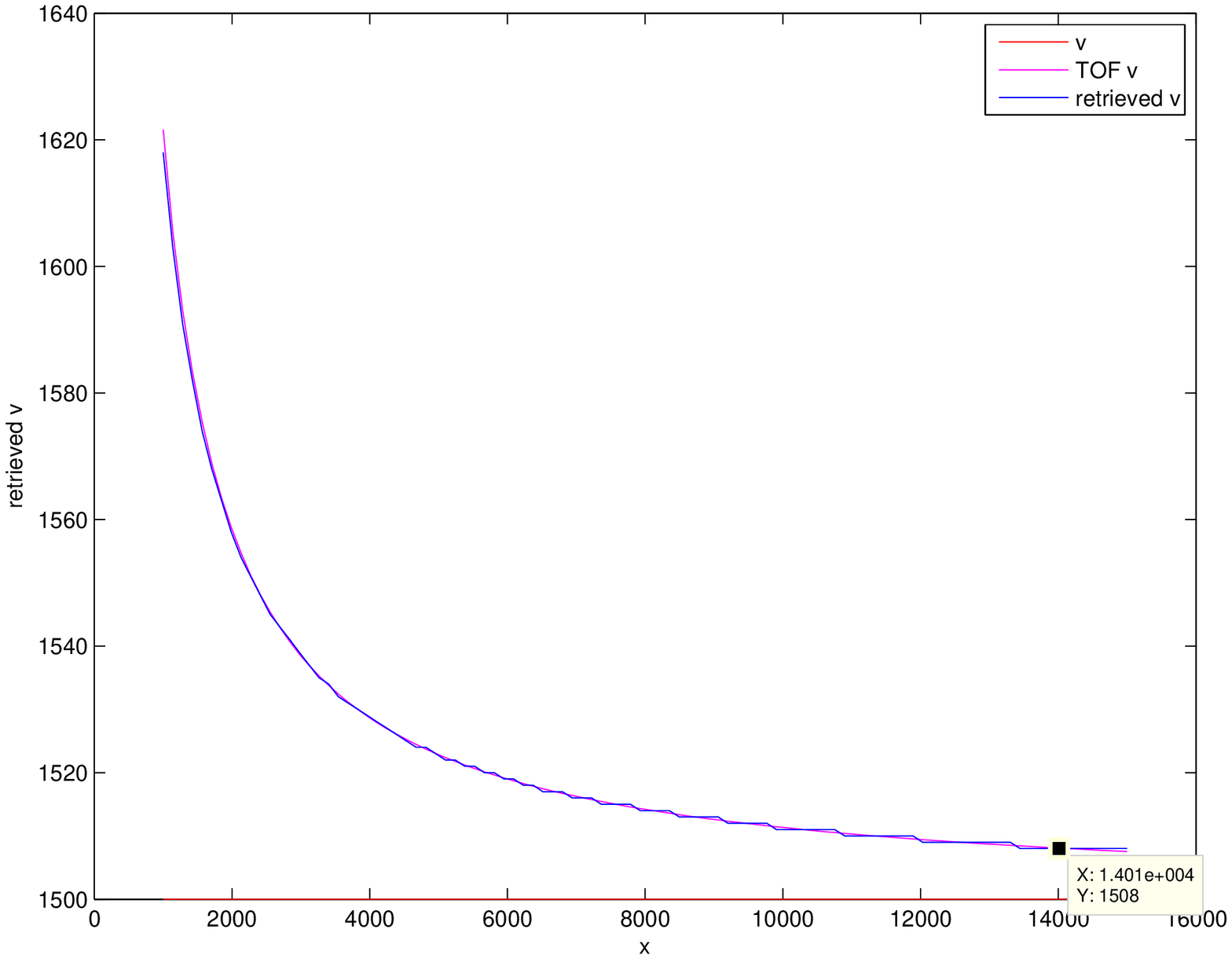}
\caption{Retrieved velocity versus $x$. The red line (labeled $v$) is the true velocity. The magenta curve  is the velocity  obtained by the TOF retrieval formula scheme. The blue curve is the retrieved velocity  $v''$ obtained by the FWI scheme. Case $\alpha=2~Hz$, $\beta=10s$, $\beta'=10.05~s$.} \label{ncts-1}
\end{center}
\end{figure}
\begin{figure}
[ptb]
\begin{center}
\includegraphics[scale=0.4]{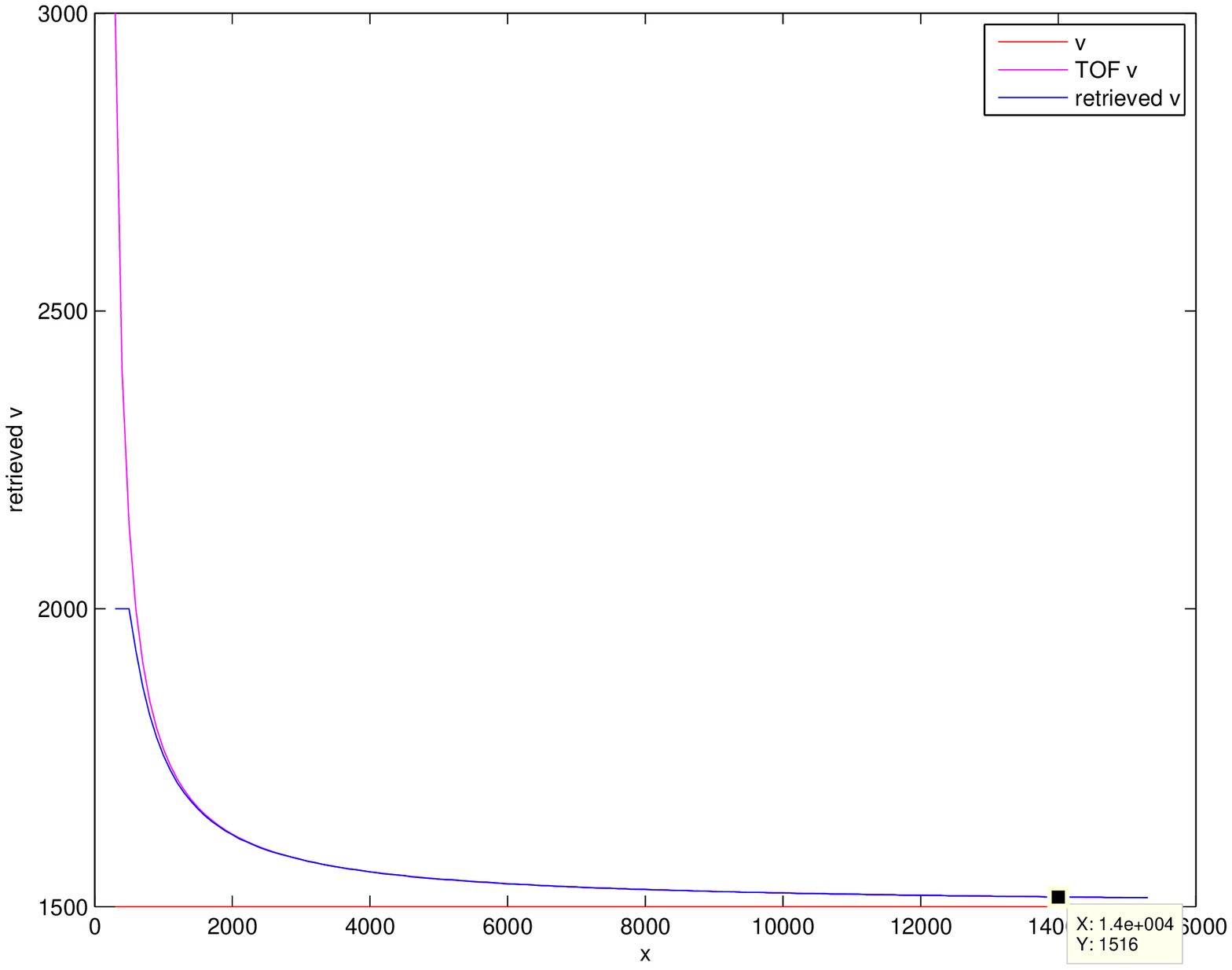}
\caption{Retrieved velocity versus $x$. The red line (labeled $v$) is the true velocity. The magenta curve  is the velocity  obtained by the TOF retrieval formula scheme. The blue curve is the retrieved velocity  $v''$ obtained by the FWI scheme. Case $\alpha=2~Hz$, $\beta=10~s$, $\beta'=10.1~s$.} \label{ncts-3}
\end{center}
\end{figure}
\begin{figure}
[ptb]
\begin{center}
\includegraphics[scale=0.4]{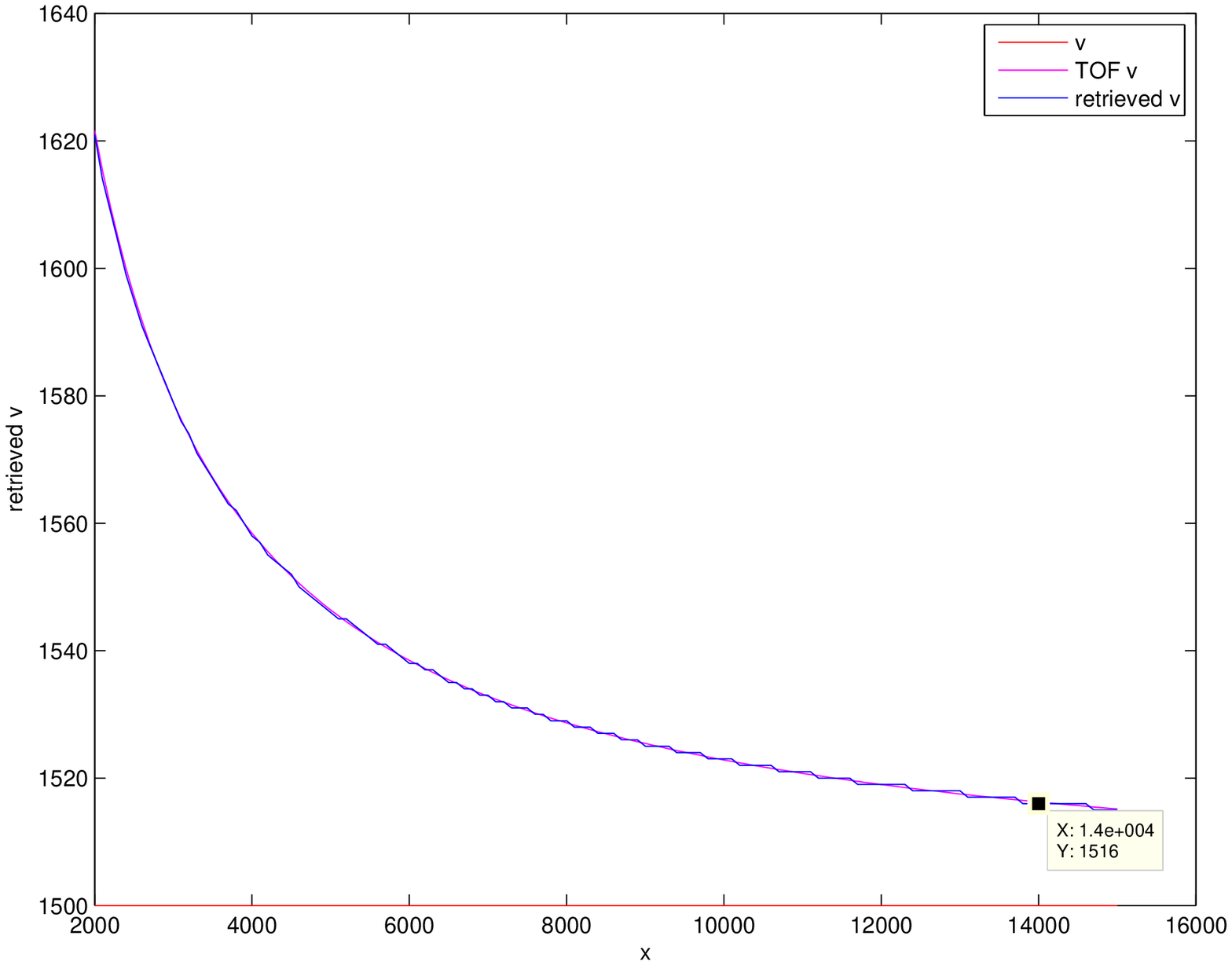}
\caption{Retrieved velocity versus $x$. The red line (labeled $v$) is the true velocity. The magenta curve  is the velocity  obtained by the TOF retrieval formula scheme. The blue curve is the retrieved velocity  $v''$ obtained by the FWI scheme. Case $\alpha=2~Hz$, $\beta=10~s$, $\beta'=10.1~s$.} \label{ncts-4}
\end{center}
\end{figure}
\begin{figure}
[ptb]
\begin{center}
\includegraphics[scale=0.4]{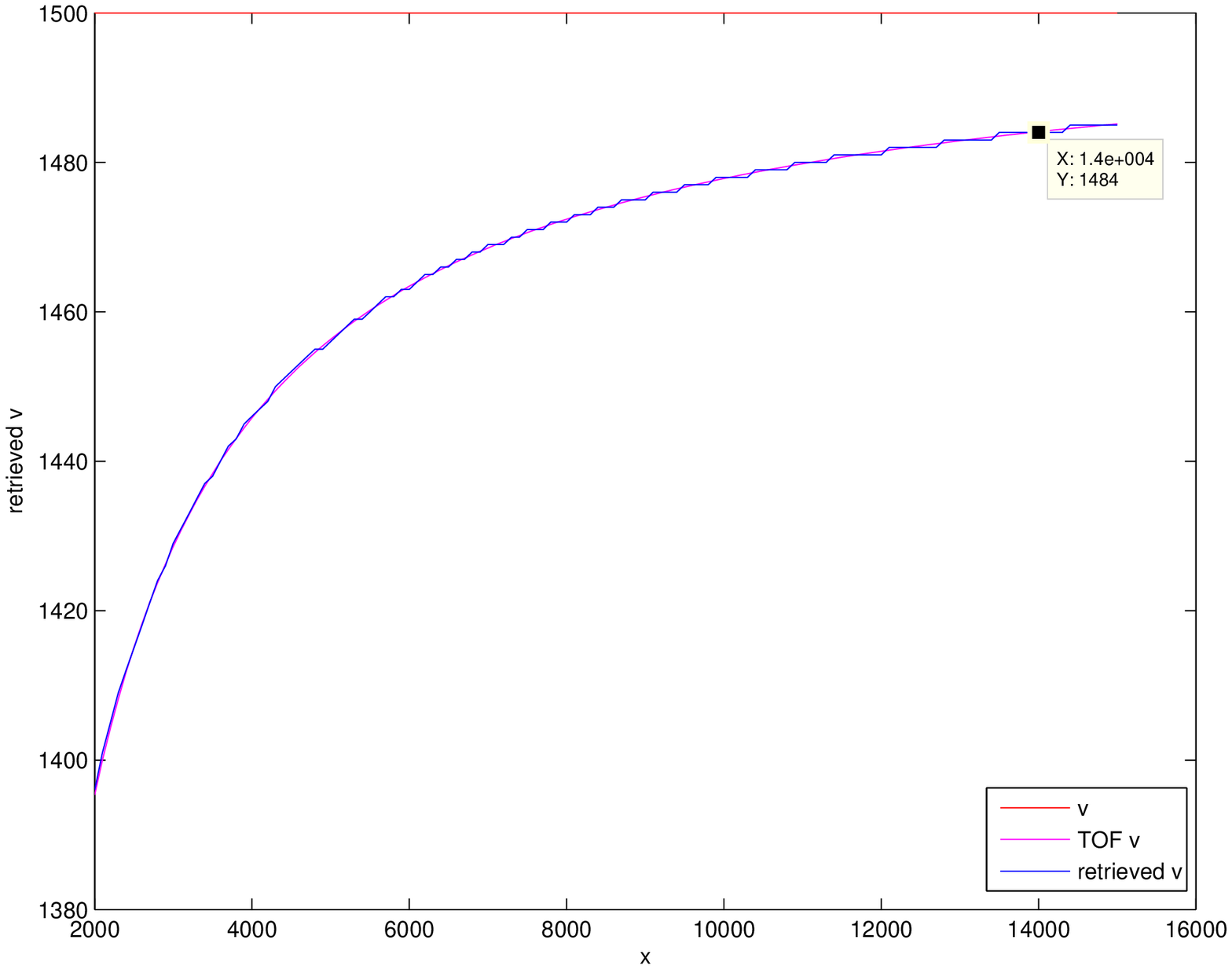}
\caption{Retrieved velocity versus $x$. The red line (labeled $v$) is the true velocity. The magenta curve  is the velocity  obtained by the TOF retrieval formula scheme. The blue curve is the retrieved velocity  $v''$ obtained by the FWI scheme. Case $\alpha=2~Hz$, $\beta=10~s$, $\beta'=9.9~s$.} \label{ncts-5}
\end{center}
\end{figure}
\begin{figure}
[ptb]
\begin{center}
\includegraphics[scale=0.4]{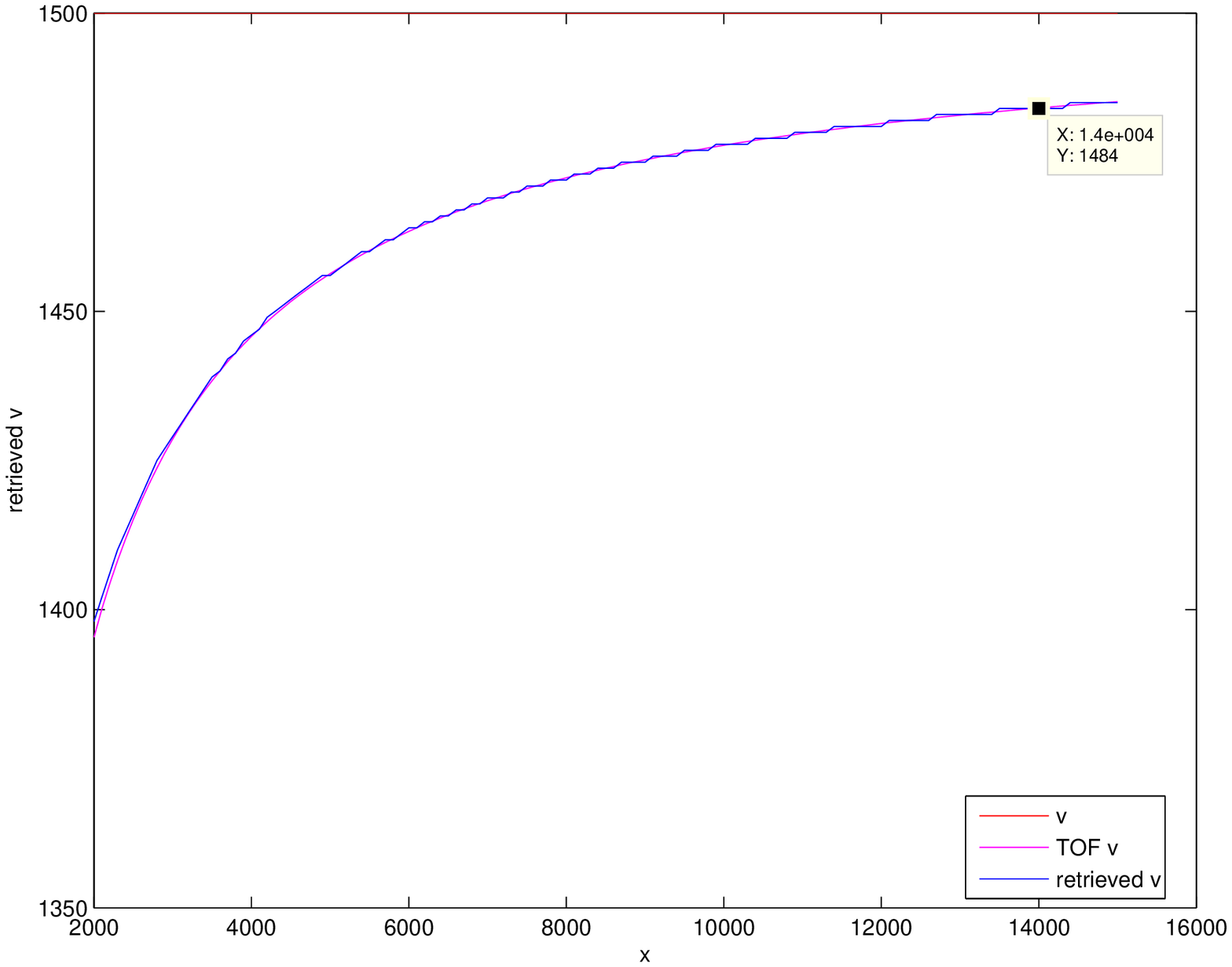}
\caption{Retrieved velocity versus $x$. The red line (labeled $v$) is the true velocity. The magenta curve  is the velocity  obtained by the TOF retrieval formula scheme. The blue curve is the retrieved velocity  $v''$ obtained by the FWI scheme. Case $\alpha=1~Hz$, $\beta=10~s$, $\beta'=9.9~s$.} \label{ncts-6}
\end{center}
\end{figure}
\begin{figure}
[ptb]
\begin{center}
\includegraphics[scale=0.4]{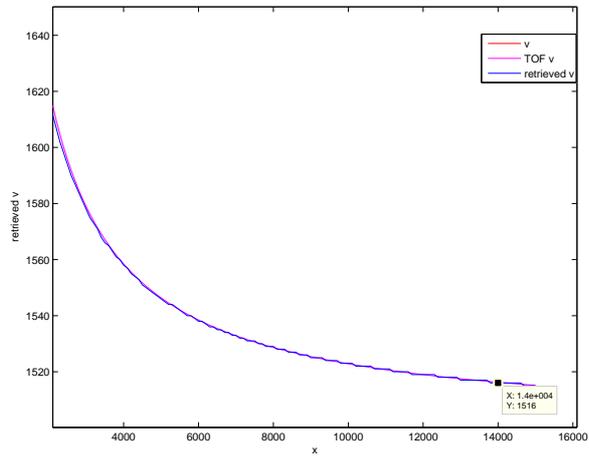}
\caption{Retrieved velocity versus $x$. The red line (labeled $v$) is the true velocity. The magenta curve  is the velocity  obtained by the TOF retrieval formula scheme. The blue curve is the retrieved velocity  $v''$ obtained by the FWI scheme. Case $\alpha=1~Hz$, $\beta=10~s$, $\beta'=10.1~s$.} \label{ncts-7}
\end{center}
\end{figure}
\clearpage
\newpage
Several features emerge from these curves:
\begin{enumerate}
\item the numerical results confirm the TOF theoretical prediction that uncertainty of $\beta$ indeed leads to error in the retrieval of $v$, with it even being {\it possible to obtain a retrieved $v$ that is larger than the true $v$}; in fact the numerically-retrieved velocity is larger (smaller) than the true velocity $v$ for $\beta'>\beta$ ($\beta'<\beta$ respectively), as predicted by the theoretical TOF analysis,
\item the two schemes (TOF and FWI) give the same results for large $x$, so that the TOF scheme can be used without hesitation for the retrieval of $v$ far enough from the source (although one has to go farther away the larger is the uncertainty of $\beta$),
\item the retrievals are not affected by the choice of pulse width (controlled by $\alpha^{-1}$), at least far from the source and when there is no uncertainty concerning $\alpha$, in agreement with the theoretical TOF analysis,
\item whatever the uncertainty of $\beta$, the farther the receiver is from the source, the smaller is the retrieval error, a fact also predicted by the theoretical TOF analysis, and which explains why it is important to acquire signal data as far as possible from the source in order to retrieve $v$.
\end{enumerate}
 The order of magnitudes of the relative retrieval errors as a function of the relative uncertainty of $\beta$ can be obtained from the following numerical results at $x=14000m$ :
\begin{equation}\label{upffa6}
\varepsilon=\frac{v'-v}{v}=\frac{1516-1500}{1500}=0.01067 \text{~~when~~} \delta=\frac{\beta'-\beta}{\beta}=\frac{10.1-10}{10}=0.01~,
\end{equation}
\begin{equation}\label{upffa7}
\varepsilon=\frac{1508-1500}{1500}=0.005333 \text{~~when~~} \delta=\frac{10.05-10}{10}=0.005~,
\end{equation}
\begin{equation}\label{upffa8}
\varepsilon=\frac{1484-1500}{1500}=-0.01067 \text{~~when~~} \delta=\frac{9.9-10}{10}=-0.01~,
\end{equation}
This shows that the relative error of $v$ is nearly the same as the relative uncertainty of $\beta$.
%%%%%%%%%%%%%%%%%%%%%%%%%%%%%%%%%%%%%%%%%%%%%%%%%%%%%%%%%%%%%%%%%%%%%%%%%%%%%%%%%%%%%%%%%%%%%%%%%%%%%%%%%%%%%%%%%%%%%%%%
\subsection{More accurate schemes for the retrieval of $v$}\label{fw}
The retrievals of $v$ observed in sect. \ref{mrvt0r} are  unreliable in that; i) they employ expressions of the field which depend on the satisfaction of the far-field and paraxial region constraints, and ii) the TOF method of retrieval is not generally applicable unless the field can be approximated by the far-field paraxial region expression.  The way to avoid these problems  is to discard the TOF scheme in favor of a {\it full-wave inversion} (FWI) method appealing to a data simulation model (when the data is simulated) and a retrieval model whose ingredients do not rely  on far-field and paraxial approximations so as to be able to account
accurately for all the physical and geometric factors that  contribute to the measured or synthetic data.
%%%%%%%%%%%%%%%%%%%%%%%%%%%%%%%%%%%%%%%%%%%%%%%%%%%%%%%%%%%%%%%%%%%%%%%%%%%%%%%%%%%%%
\section{Conclusion}
 It may be asked why  Colladon chose such a large distance (14km) between the source and the receiver. He estimated that the absolute error in observing $\Delta t$ was $0.5~s$. This is 0.056 of $9~s$ (the time it took sound to travel $14000~m$), but 0.25 of $\sim 2~s$ (i.e., the time it took for sound to travel $2500~m$). This  means that the relative error on observing  $\Delta t$  is about five times larger for the shorter distance than for the larger distance (assuming that the absolute error on $\Delta t$ is a half second in both cases, i.e., independent of the distance). But this choice  of larger $\Delta x$ to obtain a smaller relative error on $\Delta t$ was risky because the probability of the sound encountering bottom obstacles, inhomogeneities, temperature variations in the water of a natural environment are all the greater, the longer the distance traveled by the sound.

 Colladon was unaware of the fact that long distances and paraxial conditions are, in fact necessary for him to have been able to employ the TOF scheme on which these conditions are based. Nor did he know  that ignorance of the  shape of his pulse  is permissible if the conditions for employing the TOF scheme are satisfied. Above all, Colladon did not dispose of a mathematical model of his experiment, which is what we proposed to provide herein.

  This model would have told him that his TOF method can only be applied in specific circumstances and that if the latter are not satisfied, the TOF measurement result is in error. He could have saved some trouble (and further sources of error due to the variability of the physical properties of a large-scale medium) by choosing a smaller distance between the source and receiver and treating the inverse problem via the FWI (which also enables the evaluation of the error on $v$ due to uncertainty on the onset time of the pulse  as well as on other experimental parameter uncertainties.

  Nevertheless, the Colladon experiments on the kinematic method for the measurement of the velocity of sound were very ingenious (all the more so than they were carried out without the help of any electronic or even electrical devices; more recent experimental methods for such measurements can be found in publications such as (Papadakis,1967; Del Grosso \& Mader, 1972))).

  We chose not to analyze or comment on the the other aspects of the publications of Colladon \& Sturn because they speak for themselves and deal with matters that are outside  the scope of our contribution which concerned the first method (in the terminology of our Introduction) for the measurement of the velocity of sound. We hope that our translation of the material concerning the second (physical principles) method (as well as that of the first method) will be of use to those interested in the history of science, particularly those aspects of which deal with  the intense scientific activity centered on the mechanics  and thermodynamics of deformable media that took place between the seventeenth and nineteenth centuries.

%%%%%%%%%%%%%%%%%%%%%%%%%%%%%%%%%%%%%%%%%%%%%%%%%%%%%%%%%%%%%%%%%%%%%%%%%%%%%%%%%%%%%

%%%%%%%%%%%%%%%%%%%%%%%%%%%%%%%%%%%%%%%%%%%%%%%%%%%%%%%%%%%%%%%%%%%%%%%%%%%%%%%%%%%%%%%
\end{document}